%% file: main.tex
\documentclass[]{ieeeaccess} 
\input{settings}

\begin{document}

\ifdraft{
\newpage
\tableofcontents
\newpage
\listoffixmes
}{}

\doi{10.1109/ACCESS.2021.3136889}


\title{Confidential Machine Learning Computation in Untrusted Environments: A Systems Security Perspective}

\author{\uppercase{Kha Dinh Duy}, 
\uppercase{Taehyun Noh, 
          Siwon Huh,
          and Hojoon Lee} \vspace{0.5em}}

\address{\footnotesize{\texttt{\{khadinh, dove0255, calvin0420, hojoon.lee\}@skku.edu}}}

\address{Department of Computer Science and Engineering, Sungkyunkwan University}
\address{Natural Sciences Campus,  Suwon-Si, Gyeonggi-Do, Korea} 

\tfootnote{}

 
\corresp{Corresponding author: Hojoon Lee (e-mail: hojoon.lee@skku.edu).}

\begin{abstract}

As machine learning (ML) technologies and applications are rapidly changing many computing domains, security issues associated with ML are also emerging. In the domain of systems security, many endeavors have been made to ensure ML model and data confidentiality. ML computations are often inevitably performed in untrusted environments and entail complex multi-party security requirements. Hence, researchers have leveraged the Trusted Execution Environments (TEEs) to build confidential ML computation systems. We conduct a systematic and comprehensive survey by classifying attack vectors and mitigation in confidential ML computation in untrusted environments, analyzing the complex security requirements in multi-party scenarios, and summarizing engineering challenges in confidential ML implementation. Lastly, we suggest future research directions based on our study.

\end{abstract}

\begin{keywords}
confidential machine learning computation, trusted execution, side-channel attacks, multi-party ml computation
\end{keywords}

\titlepgskip=-15pt

\maketitle
\section{Introduction}

The recent advancements in \gls{ml} and its applications are bringing far-reaching changes across many fields in computing. Many areas of studies made endeavors to improve \gls{ml} in various aspects or adopted \gls{ml} for many purposes. However, security issues are posing a formidable threat to \gls{ml}-based technologies and services. For instance, the robustness of \gls{ml} models and derived services against malicious actors is under discussion in the context of adversarial machine learning~\cite{lowd2005adversarial,dalvi2004adversarial,kurakin2017adversarial,huang2011adversarial,papernot2016transferability, brendel2018decisionbased}. Also, existing research has shown that the data used during training can also be leaked through inference results~\cite{membership-inference,li2013membership,li2021membership,shokri2017membership}. Protecting \gls{ml} models and data has been the most important security objective in \gls{ml} computations, among others. The data used for training \gls{ml} models often include privacy-sensitive information in a large volume. Hence, failure to maintain the confidentiality of the data can face catastrophic consequences~\cite{holmes533}. In the case of \gls{ml} models, they are often the intellectual properties of service providers.

Secure computation of machine learning workloads has also been a main topic of interest for many systems security researchers for many years. \hl{A plethora of works in the realm of systems security have explored the security risks and solutions in protecting the \emph{confidentiality} of the protected assets such as ML models, programs, and data during computation}~\cite{ohrimenko2016oblivious,hunt2018chiron,hunt2016ryoan,quoc2020securetf,hunt2020telekine}. Confidential \gls{ml} computation is the most prominent approach in such endeavors. Confidential \gls{ml} computation adapts \glspl{tee}~\cite{sgx,tz} to protect the confidentiality of the protected assets such as the data, \gls{ml} model as well as computations performed on them. TEEs are hardware-supported technologies that can be leveraged to protect sensitive code and data. TEEs have long been adopted in implementing confidential large data computations~\cite{schuster2015vc3,priebe2018enclavedb,dinh2015m2r,zheng2017opaque}. More recently\hl{,} a number of researches \hl{have proposed} unique challenges in TEE-based confidential \gls{ml} computation~\cite{lee2019occlumency,mo2020darknetz,gangal2020hybridtee,kunkel2019tensorscone,quoc2020securetf}. 

\hl{We provide a systematic and in-depth review of the current state of the confidential ML computation through our comprehensive review of the existing works. }We identified largely three prominent topics in confidential \gls{ml} computation that the accumulation of contributions has shaped. We classify and analyze the existing works in each of the three topics: \hl{threats and} mitigation of attack vectors in untrusted computation environment\hl{s}, multi-party confidential \gls{ml} computation, and retrofitting software and hardware architecture for confidential \gls{ml} computations. 

Regarding the first topic, we analyze the security risks in terms of attack vectors, and analyze their ramifications on the confidentiality of \gls{ml} computations~(\cref{sec:in-system}). The necessity for performing the computation in an untrusted environment is common for \gls{ml} computations. The widespread use of cloud computing resources for data-intensive \gls{ml} computations is the most common example. In many scenarios, \gls{ml} computations are often performed in \emph{untrusted} environments ranging from the cloud, edge devices, and end-point devices, hence requiring protection from TEEs. However, applying TEE to sensitive computations is not without its security risks. TEE-protected workloads in untrusted environments such as the cloud face a large attack surface. Many works have shown that the confidentiality of TEE can be compromised through various side-channel attacks~\cite{wei2020leaky,moghimi2017cachezoom,brasser2017software,yan2020cache,zhu2021hermes}. There are also attacks that specifically target TEE-based \gls{ml} computations~\cite{yan2020cache,tramer2016stealing,he2021stealing,chen2021stealing,wang2019stealing,duddu2019stealing}. We also discuss vulnerabilities that might compromise confidential ML computations on the edge~(\cref{subsec:tz-side-channel}).

Secondly, we review the existing works in the context of the multi-party scenarios and their security requirements (\cref{sec:multi-party}). \hl{Confidential ML computations often must satisfy security requirements for mutually distrusting parties simultaneously}~\cite{hunt2016ryoan,hunt2018chiron,ohrimenko2016oblivious}. The participating entities of the computation, such as \gls{ml} model owner, data contributor, computation platform provider, and service user, are not necessarily the same party. Furthermore, their interests and security risks are frequently at odds, presenting unique security requirements which a confidential computation \gls{ml} scheme has to satisfy. Such \emph{multi-party confidential \gls{ml} schemes} often require more than the simple application of TEEs; rather, a comprehensive framework specific for each scenario must be designed with careful security requirement analysis~\cite{ohrimenko2016oblivious, hunt2018chiron,ozga2021perun, hanzlik2021mlcapsule,zhang2021shufflefl}.

Third, we discuss the engineering efforts that resolved the implementation challenges in building confidential \gls{ml} computation~(\cref{sec:engineering-challenges}). Resolving engineering challenges in confidential \gls{ml} computation designs and implementations have been discussed in many works. For instance, \gls{ml} computations on large data had to be split into smaller batch sizes to overcome the limited memory capacity of TEEs~\cite{lee2019occlumency, gu2020confidential}. Additionally, porting \gls{ml} frameworks (e.g., Tensorflow and PyTorch) to use TEEs is a daunting challenge, considering the complexity of such software.

This survey reviews over 140 works that contributed to building secure systems for \gls{ml} and similar data computation, focusing on the two most prevalent TEEs, Intel SGX for cloud computing and ARM TrustZone for edge computing. We summarize the contributions and insights of this paper as the following:

\begin{itemize}
\setlength\itemsep{0.3em}
\item We conduct a thorough survey of attack vectors in confidential \gls{ml} computation in an untrusted computation platform. 

\item We categorize the attack vectors on confidential \gls{ml} computation through our study of existing attacks. We also discuss available mitigations. 

\item We generalize the TEE-based multi-party \gls{ml} computation schemes into four scenarios and identify per-party security requirements. Also, we explain the contributions of the existing works according to our generalization. 


\item We summarize the engineering efforts that seek to optimize software and hardware for secure \gls{ml} computations.

\item Based on our thorough survey, we point out the relatively underexplored topics.

\end{itemize}

\input{sections/background}

\input{sections/taxonomy}
\input{sections/in-system}

\input{sections/multi-party}

\input{sections/engi-challenge}

\input{sections/opportunities}

\section{Conclusion}
This survey provides a comprehensive study of the security and engineering challenges in implementing various types of confidential \gls{ml} computation. We systematically summarized and categorized the existing works on advancing confidential \gls{ml} computation in several aspects; we discussed known in-system attack vectors and proposed mitigations, solutions for each multi-party \gls{ml} scenario, and engineering challenges in implementing secure and confidential \gls{ml} computation. Lastly, we presented our view on research opportunities based on the literature that we reviewed. We hope that our survey can serve as a comprehensive overview to systems security researchers and industr\hl{ies} who seek to build confidential \gls{ml} computation systems.
\label{sec:conclusion}

\section{Acknowledgement}
This work supported by the National Research Foundation of Korea (NRF) grant funded by the Korea government ()(NRF-2020R1C1C1011980), and Institute for Information \& communication Technology Promotion (IITP) grant funded by the Korea government (MSIT) (No. 2019-0-01343, Regional strategic industry convergence security core talent training business)

\bibliographystyle{IEEEtran}
\balance
\bibliography{IEEEabrv,references.bib}

\newpage
\begin{IEEEbiography}[{\includegraphics[width=1in,height=1.25in,clip,keepaspectratio]{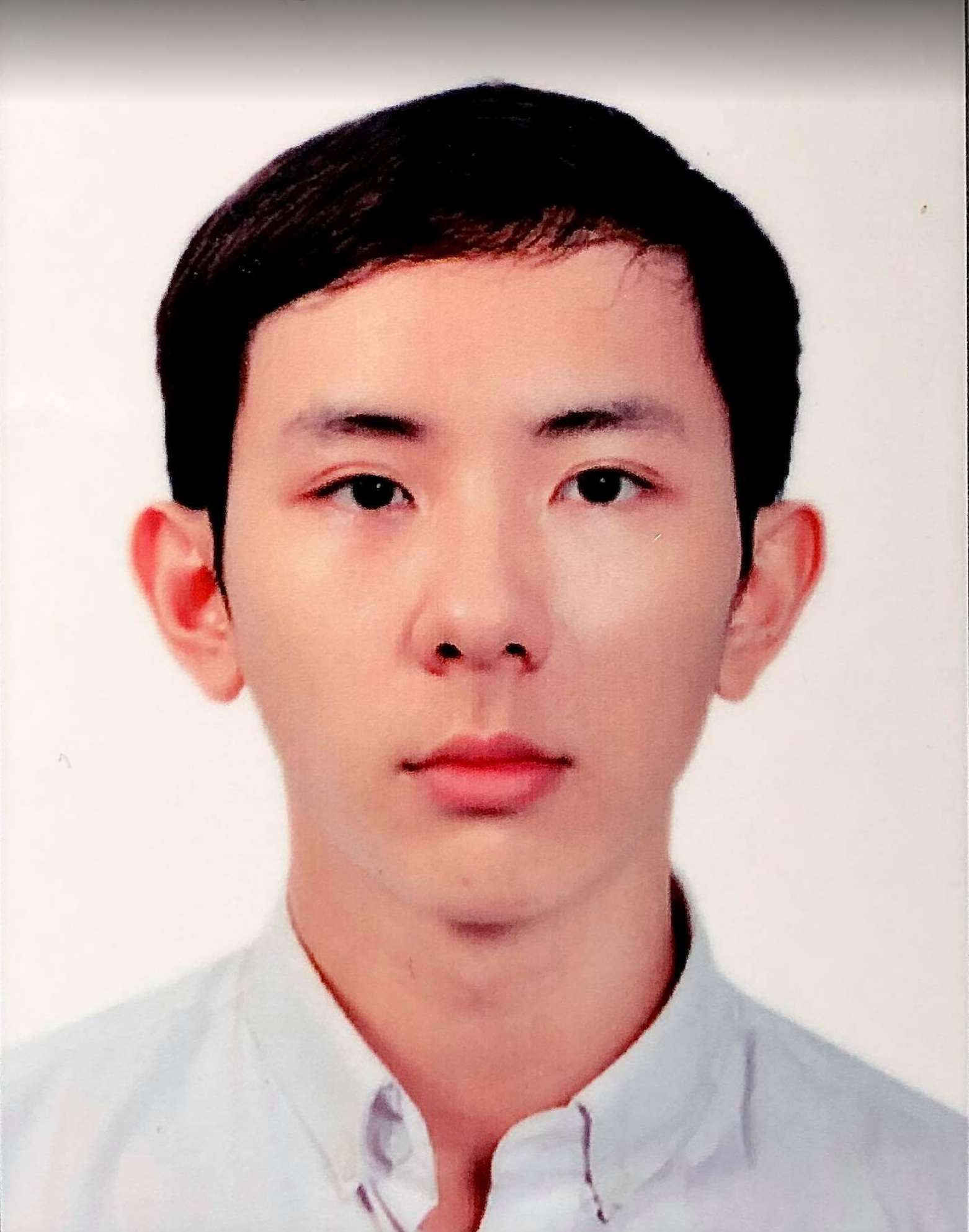}}]{Kha Dinh Duy} 
Kha Dinh is currently an integrated MS-PhD student at the Dept. of Computer Science and Engineering at Sungkyunkwan University, South Korea. He received his B.S. in Computer Science from the Hochiminh University of Technology in 2018. His main research interests are trusted execution environments, hardware-assisted security, and designing secure systems.
\end{IEEEbiography}

\begin{IEEEbiography}[{\includegraphics[width=1in,height=1.25in,clip,keepaspectratio]{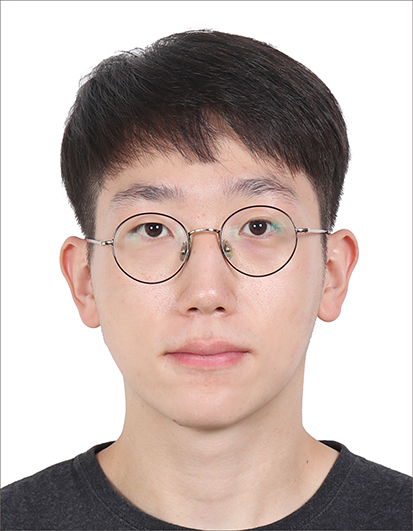}}]{Taehyun Noh}
Taehyun Noh was born in Daegu, Republic of Korea on the 1st of August 1996. He received a bachelor of software engineering from Sungkyunkwna University in 2021. He is currently an M.S. student at the Dept. of Computer Science and Engineering at Sungkyunkwan University, South Korea. At the system security laboratory, he is working on intra-process isolation technique under advisement of Professor Hojoon Lee. His research interests are virtualization techniques, secure accelerator computation and WebAssembly ecosystem.
\end{IEEEbiography}

\begin{IEEEbiography}[{\includegraphics[width=1in,height=1.25in,clip,keepaspectratio]{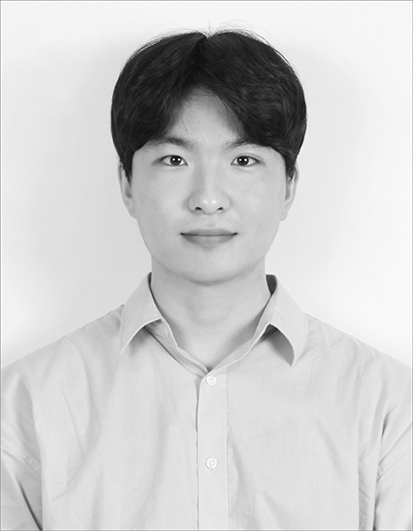}}]{SIWON HUH} received the B.S. degree in mathematics and computer science and engineering from Sungkyunkwan University in 2021. He is currently pursuing a master’s degree in computer science and engineering, Sungkyunkwan University, Suwon, South Korea. His research interests include blockchain identity management and artificial intelligence security. 
\end{IEEEbiography}

\begin{IEEEbiography}[{\includegraphics[width=1in,height=1.25in,clip,keepaspectratio]{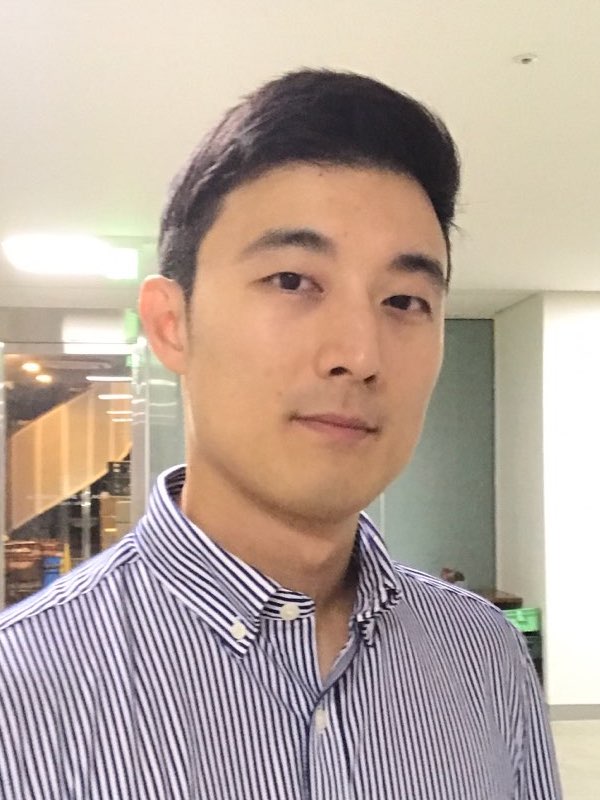}}]{HOJOON LEE} 
Prof. Hojoon Lee is currently an assistant professor at the Dept. of Computer Science and Engineering at Sungkyunkwan University since September 2019. Prior to his current position, he spent one year as a postdoctoral researcher at CISPA under the supervision of Prof. Michael Backes. He received his Ph.D. from KAIST in 2018, advised by Prof. Brent Byunghoon Kang and his B.S. from The University of Texas at Austin. His main research interests lie in retrofitting security in computing systems against today’s advanced threats. His research interests include but are not limited to Operating System Security, Trusted Execution Environments, Program Analysis, Software Security, and Secure AI Computation in Cloud.
\end{IEEEbiography}

\EOD

\end{document}

%% file: sections/background.tex
\section{Background and Related Surveys}
\label{sec:overview}
\hl{In this section, we explain the motivation for confidential ML computation that is commonly shared among the literature that we cover. Also, we describe the key techniques and terminologies that are essential for comprehending the contributions of the works that we review. Lastly, we explain the scope of the survey and its uniqueness with respect to existing surveys on the security of ML.}

\subsection{Confidentiality in ML computation}
Confidential \gls{ml} computation seeks to protect the confidentiality of the assets involved in \gls{ml} computations in untrusted computing environments.  The protected assets can include data, \gls{ml} model\hl{(s)}, \gls{ml} programs, and by-products of the computation that can indirectly undermine the confidentiality of the protected assets. In \gls{ml} computation, it may be desirable to maintain the confidentiality of the data used for training, the program that is used for training, or the resulting \gls{ml} model. \Gls{ml} computations that can be subject to protection for confidentiality are \emph{model training} and \emph{model inference}. 

\hdr{Need for confidential ML computation.} 
\gls{ml} computations often inevitably take place in untrusted environments for several reasons. First, \gls{ml} computations are often performed in the public cloud services due to their convenience. \hl{Utilizing the cloud computing services is cost-efficient compared to in-house computation infrastructures in many cases.} Second, the multiple parties may contribute to an ML process~\cite{ohrimenko2016oblivious,hunt2018chiron,ozga2021perun}. For instance, the owner of the data and the party that trains and claims the ownership of the resulting \gls{ml} model may be different parties but still want confidentiality of their assets. Third, computation platforms can be bound to a specific location due to service quality or security policy. A service that utilizes the results of inferences using \hl{an} \gls{ml} model may be latency-sensitive~(e.g., a self-driving automobile). In such cases, the \gls{ml} model may have to relocate to the location of the service for reliability and service quality. Similarly, a security policy may restrict the data from being removed from a designated infrastructure or location (e.g., medical institution). Then, it enforces the \gls{ml} model or \gls{ml} program of a different party to be stored \hl{in} an untrusted environment (from the model owner perspective).

For the reasons above, it is clear that confidential ML computation methods are necessary for securing valuable assets of ML computation in untrusted environments such as the cloud. Commercial confidential cloud computing services~\cite{microsoftazure,googlecloudconfidential,amazonnitro} that support \gls{ml} computations represent the customer demand and an emerging market for confidentiality guarantees on their workload in the cloud.

\subsection{Trusted Execution and Confidential ML Computation}
\label{subsec:bg-tee}
\hl{Confidential computing schemes that leverage trusted execution environments have been actively explored through many works to protect data and code execution}~\cite{priebe2018enclavedb,schuster2015vc3,shaon2017sgxbigmatrix,lee2019occlumency}. While cryptographic methods (e.g., homomorphic encryption) can achieve similar security goals, many academics works and existing services adapt TEE-based schemes due to their practicality in terms of performance~\cite{schuster2015vc3}.

\hdr{CPU-support for trusted execution.} \hl{Trusted execution environments (TEEs) included in commodity processor architectures} commonly provide hardware memory protection mechanisms that \hl{enable the} isolation of in-\gls{tee} program code and data. Only when the current context is in a \emph{trusted execution mode}, the isolated code and data become accessible. Hence, the usual programming model using \glspl{tee} is to split a program into trusted and untrusted domains. The trusted domain is protected \hl{by the hardware mechanisms} and only accessed through a strictly controlled interface \hl{from the untrusted domains}. Protected code inside the TEE and trusted hardware form a \emph{Trusted Computing Base (TCB)}. Intel \hl{\emph{Software Guard Extensions (SGX)}}~\cite{sgx} and ARM \emph{TrustZone~(TZ)}~\cite{tz} are the most prevalent TEEs adopted by existing commercial devices and in many research works.

\hdr{Intel SGX.} SGX~\cite{sgx} introduces a set of hardware extensions to the x86 architecture to support an isolated memory space and execution mode called the \emph{enclaves}. Each process can create its own enclave to store its sensitive code and data. The memory pages that belong to enclaves, called \emph{Enclave Page Cache (EPC)}\hl{,} \hl{are} protected by hardware from all privileged software such as the OS kernel, hypervisor, or even the BIOS. A context can enter an enclave only through the strictly controlled interface that is defined by \emph{Enclave Calls (\texttt{ECALLS})}. SGX's security model defines only the in-enclave program and \hl{the} CPU as TCB. Due to the dominance of Intel processors in the server market, SGX is adopted for many in-cloud confidential ML computation schemes.

\hdr{ARM TrustZone\hl{.}} ARM TrustZone (TZ)~\cite{tz} is commonly used on mobile and edge devices due to the ARM processors' prevalence in the market. TZ divides the processor states into \hl{the} \emph{secure world} and \hl{the} \emph{normal world}. \hl{While SGX exclude kernel from its TCB, TZ's secure world also includes a secure kernel}. Hence, the TCB of TZ is the whole software stack in the secure world, including the trusted apps and the trusted kernel. Hardware mechanisms strongly enforce the isolation between the two worlds. Although TZ employs a more coarse-grained isolation~(two security states) compared to Intel SGX~(at each process level), recent proposals also introduce the capability to host a secure enclave to TZ without requiring hardware modifications~\cite{brasser2019sanctuary,zhao2019sectee}.

\hdr{TEE designs for RISC-V.} Apart from SGX and TZ, there are also several research proposals for other ISA. In particular, Sanctum~\cite{costan2016sanctum}, Keystone~\cite{lee2020keystone} and {CURE}~\cite{bahmani2021cure} are TEE architectures designed for RISC-V. 

\hdr{TEEs and confidential computing.} Over the years, many works leveraged TEEs for confidential computing systems. Haven~\cite{baumann2014shielding} enables two-way protection \hl{of applications} by first employing SGX hardware to protect the code and data inside the enclave, then utilizes software containers \hl{to isolate} untrusted and unmodified binaries \hl{from the host system}. Subsequence works follow the same approach for two-way isolation~\cite{arnautov2016scone,hunt2016ryoan,tsai2017graphenesgx,menetrey2021twine,goltzsche2019acctee}. {SCONE}~\cite{arnautov2016scone} enables Docker~\cite{merkeldocker} services to be protected by SGX. Ryoan~\cite{hunt2016ryoan} prevents untrusted data processing modules from leaking information by utilizing NaCl containers~\cite{yee2009native}. Finally, Graphene-SGX~\cite{tsai2017graphenesgx} provides an efficient library OS for SGX based on \hl{Graphene}~\cite{tsai2014cooperation}. There are also works that utilize TEE to protect distributed computation~(e.g., MapReduce and Spark), which requires coordination from enclaves on multiple systems~\cite{schuster2015vc3,dinh2015m2r,zheng2017opaque}. All the works mentioned above \hl{provide} a strong baseline on which secure ML systems using TEEs can be built.

\subsection{Alternative approaches to trusted execution}

Besides the hardware support for TEE, there exist techniques that serve as building blocks for confidential \gls{ml} computation. Our survey does not focus on those approaches, since they are already covered by several surveys. \hl{For instance, a survey by}~\cite{cabrero-holgueras2021sok}\hl{ reviews privacy-preserving cryptographic techniques to protect privacy in deep learning. The survey by Sagar et al.}~\cite{sagar2021confidential}\hl{ focuses on cryptographic approaches for protecting confidential ML computation on untrusted platforms. Finally, Ji et al.}~\cite{dp-survey}\hl{ provides a comprehensive review of differential privacy techniques' applications in machine learning.} Therefore, we \hl{only} explain the techniques briefly here.

\hdr{Homomorphic Encryption.}
\emph{\Gls{he}} are cryptographic schemes that allow computation to be performed on encrypted data without being decrypted. \emph{Fully homomorphic encryption~(FHE)} allows arbitrary computations on encrypted data, thereby allowing computations can be performed without breaking the confidentiality of data. However, due to high-performance overhead, many recent works have resorted to TEE as a practical alternative~\cite{schuster2015vc3}.

\hdr{Secure Multi-party Computation.}
\emph{Secure \gls{mpc}} ~\cite{multiparty-unconditional, mental-game, yao1, yao2} is a confidential computing approach that employs cryptographic protocols, which splits the computation on data shared between multiple parties in a way that no individual party can see the other party’s data. \Gls{mpc} mechanisms are supported by various cryptographic building blocks, such as secret sharing, coin tossing, oblivious transfer, zero-knowledge proofs. In this survey, we only discuss TEE-based multi-party computation and leave cryptographic multi-party computations out of scope.

\hdr{Differential Privacy.}
\emph{Differential privacy (DP)} places a constraint on the algorithms used to publish aggregate information about a statistical database, limiting the disclosure of private information of records in the database. For example, differentially private algorithms are used by some government agencies to publish demographic information or other statistical aggregates while ensuring the confidentiality of survey responses and by companies to collect information about user behavior while controlling what is visible even to internal analysts. DP is essentially an algorithmic-based data handling, thus is orthogonal to the existence of TEEs. In fact, multiple TEE-based confidential ML \hl{systems} apply DP to enhance data privacy~\cite{zhang2021citadel,hynes2018efficient}.

\subsection{Scope of this study and similar related surveys}
Our survey focuses on the systems endeavors that seek to protect the confidentiality of ML models, data, and ML programs in ML computations. We systematically analyze the in-system attack vectors that might compromise confidential ML computation protected by TEEs and their proposed mitigation methods. We also discuss the case of multi-party computation, where multifold security guarantees have to be simultaneously satisfied for parties with different interests. Furthermore, we survey the problems that one faces when implementing practical TEE-based confidential ML systems. These problems range from current memory capacity limitations to unfriendly TEE development \hl{systems}. Clearly, we do not focus on adversarial attacks~\cite{papernot2018sok,liu2018survey,biggio2018wild} which aim to harm the integrity of ML services.

To the best of our knowledge, a comprehensive survey from a \emph{system security perspective} for secure ML computation has not been conducted. 
Although some surveys that cover the non-system security aspect of machine learning have been presented recently, those surveys focus on a specific type of deployment \hl{environment}~\cite{mittal2021survey,kairouz2021advances} or a technique used to achieve security~(e.g., differential privacy~\cite{dp-survey} and cryptography~\cite{mireshghallah2020privacy,sagar2021confidential,cabrero-holgueras2021sok}). Mireshghallah et al.~\cite{mireshghallah2020privacy} discuss the privacy issues of deep learning systems, but primarily focuses on the indirect threats and algorithmic and cryptographic defenses. A survey by Meurisch et al.~\cite{meurisch2020privacypreserving} \hl{discusses} both systemic challenges and algorithmic challenges of data protection in general AI services. Our survey, \hl{on the other hand}, discusses the protection of all valuable assets and is specific to ML computation.


%% file: sections/taxonomy.tex
\section{Defining Entities and Assets in Confidential ML Computation}

\Figure[t]()[width=\textwidth]{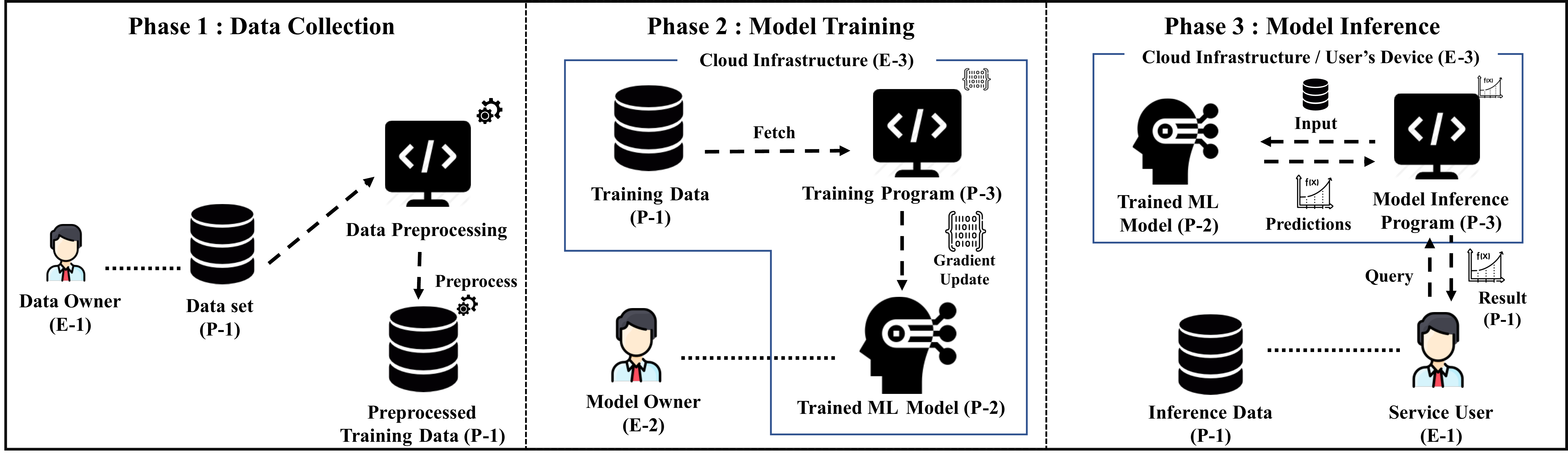}
{{An overview of the ML workflow labeled with the entities and protected assets.}\label{fig:ai-overview}}
 
Varying computation environment~(e.g., cloud vs. edge) and numerous entities involved in the computation, protected assets, and security goals complicate the security model of confidential ML computations. Hence, unified notations are essential in congregating the contributions made in each work with different attacker models and goals to paint a big picture that shows the current status of confidential ML computation. We establish a list of \hl{notations} that will be used throughout this paper such that we can explain and classify the works with consistency. \autoref{tab:taxonomy} \hl{describes the notations we used to classify the entities participating in ML computation and the protected assets processed by the ML computation.} \hl{We provide an in-depth explanation of the entities and protected assets in subsections} \cref{subsec:entities}\hl{ and} \cref{subsec:assets}.

\hdr{\hl{Overview of the ML workflow.}}\autoref{fig:ai-overview} \hl{describes an overview of the general ML workflow, annotated with our notations. Most ML computations start at the \emph{data collection phase}, where data used for training is collected into a \emph{data set}. Typically, the data set is \emph{preprocessed}, i.e., is cleaned and transformed to a form that is easier to process by the ML algorithm. Then, the computation enters the \emph{model training} phase, where a \emph{training program} containing the logic for the ML algorithm (e.g., deep neural network (DNN)) fetches batches of training data to calculate the \emph{gradient update}. The training program uses gradient updates to update the \emph{ML model} until it has processed all training data or met predetermined stopping criteria. Finally, after training finish, the trained ML model is deployed for \emph{model inference}, where the model will be employed to perform predictions. Commonly, the model is deployed in an \emph{inference service} executed in the cloud that receives users' \emph{inference data} as \emph{queries} through an API endpoint. On receiving queries, the service then feeds inference data to the trained ML model to obtain the prediction results, then send the results back to the users. As we discuss in} \cref{sec:multi-party}\hl{, there are scenarios where the model inference services must be deployed to the edge and users' devices to perform the computation.}


\label{sec:taxonomy}

\input{tables/taxonomy}

\subsection{Entities involved in ML computation}
\label{subsec:entities}

\hdr{\id{E-1}: Data Owner. } Data owners are the ones who contribute data to train an ML model (\emph{training data owners}, \id{E-1-a}) or send their private data to the model owner to request for inference (\emph{inference data owner}, \id{E-1-b}). \hl{In most cases}, data owners are users of an ML service. Data owner and model owner can be the same entity, as in the case of offloading ML computation to the cloud.

If the data owner is different from the model owner and platform provider, for instance, when a user submits their data to a cloud-deployed ML model for prediction, the data owner wishes to preserve the privacy of their data. Such privacy must be protected even when data is used by the ML model for training and inference. This means that: (1) The ML program that handles data for training or inference must respect its privacy, (2) The resulting model must not contain traces of information that trace back to the data owners or violate their privacy~(e.g., the model must be \emph{differentially private}). Both constraints have been addressed by several related works, using TEEs \cite{hunt2016ryoan,hunt2018chiron,zhang2021citadel} and using differential privacy techniques \cite{kocher1999differential}. 

\hdr{\id{E-2}: Model Owner. } Model owners are the entity that possesses the intellectual property (IP) of an ML model. In most cases, model owners want to keep their models private, as leaking the model can lead to a detrimental loss in profit for a business or introduce privacy risks. Researchers have demonstrated that sensitive information about training data can be extracted from pretrained ML models~\cite{salem2020updatesleak,fredrikson2015model}. As a result, some ML services provide only limited access to the model through API queries. 

When a model is deployed for inference, the inference data owner could be considered an adversary that wishes to steal the underlying ML model. In this scenario, the model owner also desire that no information about the model be leaked during the ML inferencing process.

\hdr{\id{E-3}: Platform Provider. } As training modern ML models often requires significant computation power, model owners and data owners often offload the computation to services that provide specialized infrastructure for ML workloads, the platform provider. In the most common case, the platform provider is a cloud service provider that is hosting the system running the computation. However, when ML services are getting deployed to the users' devices, the platform provider can also be the user of the service. 

Platform provider is commonly untrusted by parties that are using it; however, with the introduction of trusted hardware such as Intel SGX~\cite{sgx} and ARM TrustZone~\cite{tz}, the remote client of untrusted platforms can establish trusted and secure execution environments, sometimes referred to as \emph{enclaves}, on the platform. 



\subsection{Protected assets in ML computation}
\label{subsec:assets}
\addtoc{Explain about valuable assets in of machine learning, why they are valuable/ need to be confidential}

\hdr{\id{P-1}: Data. } Data is at the center of building and servicing \gls{ml} models; data defines the model behavior and the quality of the trained model. In \gls{ml} training, data is usually aggregated into a \emph{data set}, a collection of related data used for a specific purpose.  There are two types of data utilized in an ML workflow, \emph{training data}, data employed to optimize the model and \emph{inference data}, data used to obtain the prediction results and never affect \hl{the} trained model. 

Data used for \gls{ml} training and inferencing often contains privacy-sensitive information that \hl{is} directly or indirectly relates to thousands or even millions of individuals. Securing the confidentiality of the data is becoming an important ethical responsibility for research institutions and corporates as their data expands in terms of spectrum and depth, and also failure to secure the confidentiality of the data may result in costly lawsuits.

\hdr{\id{P-2}: \gls{ml} Models.}
An \gls{ml} model commonly \hl{consists} of \emph{model architecture} and \emph{model parameters}. Model architecture refers to the metadata that is associated with the topology of an ML model (e.g., the layers of the neural network, the connection between each layer, or the activation functions, etc.).  Model parameters, sometimes referred to as model \emph{weights}, are tuned as a result of model training. The model parameters and architecture are typically stored separately or together by \gls{ml} frameworks in a compact binary format such as Protocol Buffers~\cite{googleprotocol} or HDF5~\cite{folk2011overview} that can be efficiently loaded by the applications later for inference. Efforts also have been made to make model files more portable across platforms through the ONNX open format~\cite{thelinuxfoundationonnx}. 

Developing and training mature \gls{ml} models often require a tremendous amount of effort, and therefore considered intellectual property to be protected unless specifically declared to be public. It would require a well-thoughtout data modeling plan and a possibly long period of data acquisition. Designing \hl{and training} a \gls{ml} model \hl{would} require highly trained data scientists, not to mention the cost of infrastructure for computing  (e.g., cloud computing resources). 




\hdr{\id{P-3}: \gls{ml} Program. }
The \gls{ml} program describes the procedures to handle ML models and data, and the procedures in which the ML model \hl{interacts} with data. 
Model hyperparameters are typically defined by \hl{the} \gls{ml} program instead of residing in a model file. Hyperparameters refer to information of the model associated with the learning process, which includes the learning rate, batch size, and regularization factors. 
Since the program also contains the trade secrets of an ML service provider, such as the method of processing data or how user's input is handled, they also need to be kept private.

%% file: tables/taxonomy.tex
\begin{table*}[t]
    \centering
    \begin{tabularx}{0.95\textwidth}{ccX}
        \toprule
        \thead{ID} & \thead{Entity} &\thead{Description}  \\        
        \midrule
        \textbf{\texttt{E-1}} & Data Owner & The one who contributes data to train a model (\id{E-1-a}), or the user who sends private data to an ML service for inference (\id{E-1-b})\\
        \textbf{\texttt{E-2}} & ML Model/Program Owner & Model owner who desires confidentiality of her model and program \\     
        \textbf{\texttt{E-3}} & Computation Platform Owner & The entity who has full control over the device in which ML computation is performed (e.g., cloud service provider, device owner) \\        
        \toprule
        \thead{ID} & \thead{Protected Asset} &\thead{Description}  \\        
        \midrule
        \textbf{\texttt{P-1}} & Data & Data or dataset used for training a ML model or use for inferencing\\        
        \textbf{\texttt{P-2}} & ML Model & Trained Machine Learning Models\\     
        \textbf{\texttt{P-3}} & ML Program & Program that instructs training procedure to yield ML model \\

        \bottomrule
         
    \end{tabularx}
    \caption{\hl{The notations (\textbf{ID}) for} entities, protected assets and computation performed in confidential ML computation \hl{and their description}. \hl{We explain existing literature and confidential ML scenarios in terms of the shown notation to maintain consistency and enable comparisons of existing works.} }
    \label{tab:taxonomy}
\end{table*}

%% file: sections/in-system.tex
\section{Securing Offloaded ML Computations in The Cloud}

\label{sec:in-system}
\input{tables/attack-vectors}

Data-intensive ML computations are becoming one of the most common workloads in cloud computing services. Due to Intel's dominance in the cloud computing market, a plethora of works has leveraged Intel SGX~\cite{sgx} to achieve confidential computation in the cloud~\cite{dinh2015m2r, priebe2020sgxlkl, lee2019occlumency,hunt2018chiron,hunt2016ryoan}. Microsoft's Azure offers SGX-enabled computing resources~\cite{microsoftazure}.

\hl{Many} works that discuss building secure ML computation in \hl{the} cloud seek to protect at least one of the following: Data (\id{P-1)}, ML model  (\id{P-2)}, and ML program  (\id{P-3)} offloaded by the client. Also, many works assume that a single entity has ownership (\id{E-1}, \id{E-2}, \id{E-3}) \hl{of} all three protected assets (we discuss multi-party computation separately in~\cref{sec:multi-party}). The offloaded workload, while protected by SGX, faces formidable adversaries. Untrusted cloud service infrastructure can launch powerful attacks with system software (e.g., kernel) privilege~\cite{bulck2017telling,vanbulck2017sgxstep} or even physical access to the hardware~\cite{membuster,murdock2020plundervolt}. An untrusted co-tenant may launch attacks that \hl{abuse} resource sharing that inevitably occurs in the cloud~\cite{yan2020cache}.

\autoref{tab:attack-vectors} \hl{outlines the reviewed offensive and defensive research works related to ML computations, categorized by the \emph{attack vectors (AV)} in untrusted environments. We group the attack vectors into three groups, attack vectors caused by untrusted software (SW), attack vectors obtained through physical access and accelerator-related attack vectors.} \hl{The following subsection (}\cref{subsec:sgx-guarantee}\hl{) discusses the basic security guarantees provided by protecting ML computations in SGX.} \hl{We describe the attack vectors that are not mitigated by SGX by design in}~\cref{subsec:attack-vectors}
\hl{.In }\cref{subsec:known-attacks}, \hl{we discuss their impact on confidential ML computation through papers that directly discuss the impact of the attack vectors in confidential ML computations. In }\cref{subsec:extend-sgx}\hl{, we discuss works that extend on the basic security guarantees of SGX to cover side-channels from ML computations and introduce the ability to offload to external accelerators. Finally, in }\cref{subsec:tz-side-channel}\hl{, we briefly cover the microarchitectural side-channels of TrustZone, the TEE of edge devices.}

\subsection{SGX Security Guarantees} 
\label{subsec:sgx-guarantee}


SGX enclaves are protected from attack vectors \id{AV-1-a} and \id{AV-2-a} by design, as explained in ~\cref{subsec:bg-tee}. The confidentiality of computations inside the enclave is preserved with hardware-enforced memory isolation from all other system execution modes~(e.g., kernel, hypervisor, or BIOS). Furthermore, SGX's Memory Encryption Engine protects the confidentiality and integrity \hl{of} the memory used by secure enclaves~\cite{costan2016intel}. Several physical attack vectors such as cold boot attack and DMA accesses (\id{AV-2-a} in \autoref{tab:attack-vectors}) are prevented by SGX hardware, as the memory encryption engine automatically encrypts and decrypts the memory accesses to the EPC region as they leave the CPU package. However, an attacker with physical access could still observe the memory access pattern.





\hdr{Remote attestation.} SGX's hardware cryptographic functions and remote attestation feature allow the remote user to verify the identity and \hl{the} offloaded program's initial state. The remote user can request \hl{proof} from the enclave and query the Intel CA to verify that the authenticity of the SGX-capable CPU in the cloud. Furthermore, the enclave sends a signed measurement result on its initial program state, thereby proving that it is indeed executing the program that the remote user had sent. These cryptographic exchanges included in the remote attestation procedure \hl{become} the basis for remote users' trust in the enclave in the untrusted cloud. In other words, through remote attestation, trust is extended from the remote user's machine to the program running in the enclave. Subsequently, the remote user could send the secret to the enclave through an encrypted channel. In the ML computation case, the secret is data, ML model, and the ML program. 

\hdr{Using SGX to secure ML computation.} An ample amount of work has used SGX to build secure ML systems by leveraging the security guarantees of SGX~(e.g., Occlumency~\cite{lee2019occlumency}, TensorScone~\cite{kunkel2019tensorscone}, secureTF~\cite{quoc2020securetf} and TF Trusted~\cite{tf-trusted}). Those works use SGX to protect the confidentiality of ML assets when the computation is offloaded to an untrusted cloud. As SGX hardware already provides most of the protection through hardware mechanisms, most works try to solve the engineer challenges associated with adapting ML computation for SGX, e.g., the memory limitation. We will further discuss the challenges that those face in more \hl{detail} in~\cref{sec:engineering-challenges}. Moreover, we cover the works that apply SGX to build secure multi-party ML systems in~\cref{sec:multi-party}. In this section, we will only cover techniques that improve upon the basic security guarantees of SGX in protecting the ML computation, namely side-channels mitigation and extending trust to external accelerators. 
    
\subsection{Threats Against Enclave Confidentiality}
\label{subsec:attack-vectors}
Protecting the confidentiality of in-enclave protection proved to be a daunting challenge. A large volume of works presented \emph{side-channel attacks} that \hl{undermine} the confidentiality of the enclaves through various attack vectors that were not considered in the initial SGX security model. \hl{Those} attacks collect execution times or access \hl{patterns of protected programs} to infer sensitive information~\cite{xu2015controlledchannel,lee2017inferring,moghimi2017cachezoom}. The existence of side-channel is becoming the most formidable challenge in providing security guarantees in ML and other general computations using SGX in the cloud. 

\subsubsection{Software Side-channel Attacks}
\hdr{Controlled channel attacks~(\id{AV-1-b}).} SGX's TCB does not include the OS kernel, yet enclaves must interact with the kernel for services such as system calls and interrupts. Researchers have exploited side-channels in such interactions, \hl{called} \emph{controlled \hl{channels}}. The controlled channel attack on SGX-protected computation \hl{was} first introduced by Xu et al.~\cite{xu2015controlledchannel}. In this attack, a malicious privileged software (i.e., hypervisors and OSes) unmaps enclave pages and monitor page faults of the enclave application to extract the access pattern. The initial controlled-channel attack can only infer \hl{a} \emph{coarse-grained} access pattern at the page level. Subsequence works leverage the kernel's ability to raise interrupts in combination with microarchitectural cache side-channels to improve upon the granularity of leaked information~\cite{moghimi2017cachezoom,vanbulck2017sgxstep}. 



\hdr{Microarchitectural side-channel attacks~(\id{AV-1-c}).}
Although applications protected by TEEs have strong confidentiality and integrity of codes and data, microarchitectural side-channel attacks~(\id{AV-1-c}) are a prevalent threat to TEE-protected computations~\cite{moghimi2017cachezoom,gotzfried2017cache,brasser2017software}. Researchers have shown that the newly discovered microarchitectural attacks on Intel x86 processors also affect SGX-protected executions. Most microarchitectural attacks \hl{aim} to extract the access pattern by leveraging side-channels branched from shared microarchitectural resources between the victim process and the attacker. Researchers have leveraged microarchitectural features such as the \hl{Translation Look-aside Buffer (TLB)} channel~\cite{gras2018translation}, last branch record (LBR) feature~\cite{lee2017inferring}, branch predictor~\cite{evtyushkin2018branchscope}, line fill buffers (LFB)~\cite{vanschaik2019ridl}, and cache~\cite{moghimi2017cachezoom,wei2020leaky,gotzfried2017cache}) to steal secret information from software protected by SGX.





\boxedpar{Adversaries with control over untrusted software within \hl{a} system can still extract memory access patterns and subprogram execution times \hl{of} TEE-protected computations \hl{through side-channels}.}

\subsubsection{Attacks with Physical Access (\id{AV-2}) }
Physical attack is another concern for the computations protected by TEEs. Multiple works demonstrate that the bus, where memory accesses and communication packets travel from the CPU to other system components, creates attack vectors that can impact the confidentiality of computation protected by SGX.

\hdr{Leaking information on the bus~(\id{AV-2-b}). } In SGX, \hl{although} enclave memory contents are automatically encrypted as it leaves the CPU package (e.g., store to DRAM), the \emph{address access pattern} side-channel is still observable by an adversary who has a snooping device set up on the memory bus. For instance, Membuster~\cite{membuster} has shown the feasibility of \hl{the} bus snooping attack; the work has shown that they could extract the queried English words \hl{from an} enclave-protected dictionary program along with other examples. Therefore, we recognize the physical as another attack vector that must be considered in confidential computing design.

\boxedpar{Adversaries with physical control over the system can leverage access to the bus to capture the memory access pattern of programs protected by TEE.}





\subsubsection{Attacks on Accelerators (\id{AV-3})} 
ML computations are often highly parallelizable and can immensely benefit from the use of GPUs~\cite{sze2017efficient,oh2004gpu} and dedicated ML accelerators~\cite{chen2017eyeriss,jouppi2017indatacenter}. However, utilizing accelerators to accelerate confidential computation introduces several issues that could compromise the security of TEEs, as the accelerator hardware itself is not included in the TCB of TEEs. 

\hdr{Insecure communication between CPU and accelerators (\id{AV-3-a}).} Add-on accelerators that communicate with the processor through an open and unencrypted medium (e.g., system bus) can leak information under powerful attacks such as physical bus snooping or communication path manipulation by the untrusted OS. As most external accelerators do not provide encryption in communication, an attacker can easily reverse-engineer the packets to infer sensitive information about offloaded workloads. We show an existing attack \hl{that} utilizes this attack vector to compromise assets of ML computation~\cite{zhu2021hermes} in the next section.


\hdr{Side-channels in sharing accelerator~(\id{AV-3-b}).} 
Side-channels are also inevitable in the \hl{accelerator itself} since the use of peripheral devices \hl{is} shared between software \hl{in most systems}. Enforcing the secure usage of accelerators~(e.g., the GPUs) would \hl{require} hardware modifications to the accelerator architecture itself~\cite{volos2018graviton} or to the host-accelerator interfaces~\cite{jang2019heterogeneous}. \hl{Some} research works \hl{exploit} side-channels from shared GPU usage to extract sensitive information from \hl{the} program executing on the GPU~\cite{dipietro2016cuda,jha2020deeppeep,wei2020leaky}.

\subsection{Known Attacks on Confidential ML Computation}
\label{subsec:known-attacks}
A number of works have devised attacks specific to ML computations to leak protected information. 

\subsubsection{\hl{Software} Side-channel Attacks \hl{on} ML Computation~}
\hdr{\hl{Cache side-channels (}\id{\hl{AV-1-c}}).}
The most common microarchitectural attacks on confidential AI discussed in the literature are those which exploit the memory access pattern obtained through cache side-channels to violate the confidentiality of ML models~(\id{P-2}) or data used in ML computation~(\id{P-1}). For instance, Privado~\cite{grover2019privado} demonstrated an attack on a secure inference service inside SGX that leaks information about the input data (\id{P-1}). In the attack, an attacker infers the output of the TEE-protected ML inference service by observing memory access patterns leaked from side-channels. More importantly, several researchers also demonstrated attacks that are able to recover information about the ML model~(\id{P-2}) just from the cache side-channels. Ganred~\cite{liu2020ganred} employs a \gls{gan} to reconstruct the target ML model from cache side-channel timing information. 

\boxedpar{Memory access \hl{patterns} collected from side-channels can be exploited to extract protected inference queries and even ML models. }

\hdr{Risks of page sharing.} Existing works have shown that an untrusted co-tenant in the cloud, a less powerful adversary compared to an untrusted cloud administrator, can still undermine the confidentiality of the ML model (\id{P-2}). Cache Telepathy~\cite{yan2020cache} exploits shared library pages that are shared among co-tenants to launch cache side-channels (\id{AV-1-c}). More specifically, it shows that system-wide sharing of physical pages that store \hl{\emph{general matrix multiplication (GEMM)} operations} can serve as a side-channel that allows the adversary to extract \hl{the} \gls{dnn} model's architecture \hl{of other co-tenants}.

\boxedpar{Sharing of resources among co-tenants in the cloud may serve as a potential side-channel for eavesdropping on ML computation.}
\subsubsection{\hl{Attacks with Physical Access and Attacks on ML Accelerators}}
\hdr{Attacks with leaked information on the bus~(\id{AV-3-a} and \id{AV-2-b}).}
Multiple research works have demonstrated that an attacker can utilize physical access to the bus to extract sensitive information from confidential ML models~(\id{P-2}). For instance, Hua et al.~\cite{hua2018reverse} demonstrate an attack that extracts \emph{\gls{cnn}} models deployed on a \gls{cnn} accelerator. The attack \hl{starts with} feeding inputs to the accelerators, \hl{observing} the memory access pattern by the accelerator on the bus\hl{,} and the timing between accesses to reconstruct the ML. On the other hand, the Hermes attack~\cite{zhu2021hermes} captures and analyzes PCI-e packets sent from the CPU and to the GPU, then \hl{uses} the \hl{obtained information} to reconstruct the entire \gls{dnn} model.

\boxedpar{\hl{The} use of accelerators through insecure I/O channel may allow \hl{an} adversary to extract protected ML models. }


\hdr{Side-channels in sharing GPU~(\id{AV-3-b}).}
Sharing a GPU in the cloud proved to be a side-channel that can be leveraged by a malicious co-tenant to extract information on a GPU workload. Works have also demonstrated that \hl{such information} can leak information of ML computation, in particular, the confidential ML model~(\id{P-2}). Several works exploit the side-channel from \emph{sharing} the GPUs to recover the \gls{dnn} architecture ~\cite{jha2020deeppeep,wei2020leaky}. Leaky \gls{dnn}~\cite{wei2020leaky} demonstrates an attack that extracts the \gls{dnn} model inside GPUs by monitoring the resource usage of the victim kernel using GPU's built-in performance counters. DeepPeep~\cite{jha2020deeppeep} combine\hl{s} multiple GPU-based side-channels~(e.g., memory footprint, timing, power usage\hl{,} and kernels percentage) to reverse-engineer the target \gls{dnn} model offloaded to the GPU.


\boxedpar{Sharing of GPU among cloud co-tenants may serve as a side-channel for information leak.}

\hdr{Power and EM side-channels~(\id{AV-2-c}).} \hl{Some} research works demonstrated that power analysis attacks~\cite{xiang2020open} and \emph{Electromagnetic (EM)} analysis attacks~\cite{yu2020deepem} \hl{might} compromise confidential ML assets. For instance, several attacks predict the model architecture and parameters observing the power consumption and EM emission~\cite{yu2020deepem,wei2018know,batina2019csi,xiang2020open}. Mitigation against those attack vectors requires careful consideration at the architectural level, which is challenging. 

\subsection{\hl{Extending SGX's security guarantees}}
\label{subsec:extend-sgx}
\subsubsection{\hl{Side-channel} Defenses for Confidential ML Computation}
\label{subsec:defenses}
\label{subsec:side-channel}
The generally accepted mitigation for side-channel leakages is \emph{input-oblivious} algorithms. Such algorithms exhibit indistinguishable memory trace regardless of input values. 

\hdr{Oblivious computing. } The most common approach is to use compiler transformations to turn \emph{input-dependent} code into input-oblivious code. Input-oblivious algorithms could prevent most of the attack that aims to leak the \emph{memory access pattern} of code and data protected by TEEs, e.g., microarchitectural side-channel attacks~(\id{AV-1-c}) and memory snooping attack~(\id{AV-2-b}). Making the entire code base of ML frameworks input-oblivious is expensive in terms of performance. For instance, Raccoon~\cite{rane2015raccoon}, a state-of-the-art method to achieve obliviousness, exhibits an average performance overhead of $21.8\times$, which makes it unsuitable for performance-sensitive ML workloads. \hl{Hence,} some works modify only the possible source of side-channel in \hl{an} ML algorithm that may leak sensitive information of data to mitigate the poor performance of \hl{the} input-oblivious algorithm~\cite{ohrimenko2016oblivious,grover2019privado}. 

\hdr{Making ML programs oblivious.} In~\cite{ohrimenko2016oblivious}, Ohrimenko et al. make ML computation oblivious at the algorithm level. The authors employ data-oblivious primitives~(e.g., oblivious assignment, comparison, array access\hl{,} and sorting) to construct oblivious ML algorithms that are free of side-channel. Five oblivious machine learning algorithms are introduced: K-Means, \gls{cnn}, \gls{svm}, matrix factorization\hl{,} and decision trees.
Privado~\cite{grover2019privado} made two key observation. First, most of \hl{the} \gls{dnn} computation\hl{s} only \hl{involve} linear operations, which are data-oblivious. Second, several types of \gls{dnn} layers have input-dependent memory access pattern\hl{s}~(e.g., ReLU and max-pool layers). The authors propose a framework that applies compiler techniques to eliminate side-channels from \gls{dnn} models. Overall, we observe that because \gls{dnn} algorithms mostly consist of input-oblivious operations~\cite{ohrimenko2016oblivious,grover2019privado}, mitigating the side-channels from those algorithms incur minimal overhead~($0.02\%$ for \gls{cnn} training in~\cite{ohrimenko2016oblivious} and $17.18\%$ on average in Privado~\cite{grover2019privado}).   

\boxedpar{
ML algorithms can be transformed such that they are input-oblivious through algorithm redesign or compiler-based techniques.
}

\subsubsection{Extending Trust to Accelerators}
\label{subsec:accelerators}
Most \gls{tee}-based confidential ML computation literature we reviewed in this survey assumes CPU-only computation. This is due to the SGX's security model that does not include OS kernel inside its trusted codebase. This means that SGX has to communicate to the accelerators such as GPU via the untrusted medium (i.e., kernel). As we will discuss \hl{in this section}, the current SGX security model requires the accelerators to actively participate in establishing a secure channel for the inclusion of acceleration in confidential computation boundaries. However, there is no known commercial-grade accelerator with such capabilities. As such, a number of works have proposed ways to utilize acceleration without compromising confidentiality or GPU architectures that include the aforementioned essential feature for confidential computation.

\label{subsec:accelerator}
\label{subsec:offload-conf}

\hdr{Blinding and verification.} As computation executed on ML accelerators is subjected to several attack vectors discussed in~\cref{subsec:attack-vectors} previously, sending plaintext data to accelerators is not secure. Moreover, the integrity of results obtained from external accelerators cannot be guaranteed due to the threat of compromised accelerators. \hl{Many works proposed techniques revolving around} a combination of \emph{blinding} and \emph{\hl{probabilistic} result verification} to allow workloads to be securely outsourced to external accelerators.

Blinding schemes \hl{\emph{obfuscate}} the offloaded workload with \hl{a random noise vector} (the \emph{blinding factor}), then \emph{de-blind} the results upon retrieval. \hl{The approach seeks to ensure the confidentiality of the offloaded tasks. } On the other hand, \hl{probabilistic verification schemes such as} the Freivalds' algorithm \hl{intend to verify the correctness of the result returned from the untrusted accelerators. The algorithm} probabilistically verify the offloaded linear operations~(e.g., matrix multiplication) correctness with reasonable accuracy and performance overhead. \hl{However, the scheme does not provide deterministic (e.g., cryptographically secure) integrity guarantees.}


A number of works apply blinding and result verification to protect the workload offloaded to accelerators~\cite{tramer2019slalom,narra2019privacypreserving,sun2021shadownet}. Out of all, Slalom~\cite{tramer2019slalom} is the first to propose the use of both method\hl{s to build a secure ML inference service} \hl{which protects the user's input when it is offloaded to an external accelerator}. Multiple following works extend on Slalom with improvements of performance and security~\cite{sun2021shadownet,narra2019privacypreserving,asvadishirehjini2020goat}. For instance, while Slalom only support\hl{s} inference workloads, GOAT~\cite{asvadishirehjini2020goat} improves the performance with hyperparameters adjustment and allows \hl{offloading training operations}. ShadowNet~\cite{sun2021shadownet}, on the other hand, improve Slalom security by also protecting the offloaded model's confidentiality against the untrusted system.

Performance is one of the biggest issues with applying blinding and verification schemes. The offloaded computation can achieve relatively higher throughput than executing the same workloads in SGX, thanks to specialized acceleration hardware. However, the blinding operation is costly both in terms of execution time and memory usage, which are resources that are scarce in TEE\hl{-}protected computations. Particularly, the blinding procedures \hl{require} expensive operations to load encrypted blinding factor\hl{s} into the enclave memory \hl{and to blind the workload}~\cite{tramer2019slalom}. Moreover, to blind a batch of data, it requires a similarly sized vector of blinding factor, which effectively halves the usable enclave memory. 

\boxedpar{\hl{Blinding schemes seek to provide confidentiality through obfuscation. Probabilistic verification schemes probabilistically ensure the integrity of offloaded computations. The two approaches provide confidentiality and integrity simultaneously, but their guarantees are not deterministic. Also, using the two techniques incurs significant performance overheads.}} 




\label{subsec:secure-accelerator}


\hdr{GPU architectures that support TEEs.} Graviton~\cite{volos2018graviton} proposes \hl{a} GPU architecture \hl{that} has built-in support for cooperative confidential computing with the host CPU enclave (e.g., SGX), thereby allowing acceleration of confidential workloads. The architecture introduces lightweight modifications to the command processor of the baseline GPU, with functionalities to support remote attestation and secure context management. The authors also introduce mechanisms to securely isolate between individual GPU contexts through protected memory regions and the ability to securely manage its own address space. The \hl{prototype was implemented using software changes that emulate the ideal hardware changes, and} \hl{shows limited performance overhead, only $17-33\%$ compared to that of native GPU execution}. \hl{Most of the reported} overheads come from data encryption traffic between CPU and GPU. \hl{Also,} the \hl{TEE-aware GPU hardware can support} a remote client to \hl{securely} utilize GPU computation on an untrusted cloud even without trusted processors such as SGX. \hl{For instance,} in Telekine~\cite{hunt2020telekine}, Hunt et al. proposed an end-to-end system that allows remote users to construct a secure communication channel to the TEE-enabled GPU using \emph{API remoting}. In the system, the untrusted server that hosts the TEE-enabled GPU only plays the role of relaying communication. Careful considerations are also made to mitigate side-channels from the timing differences. 

The approaches presented in these works indicate that the cooperation of the accelerators is an essential element in securing offloaded ML computation. The accelerator itself must be equipped with capabilities required for secure communication, including secure key storage, secure firmware that can perform self-integrity checking, bidirectional remote attestation, and symmetric key channel encryption support, etc. The accelerator's user can establish a secure communication channel between CPU and accelerators to thwart any eavesdropping attackers (\id{AV-3-a}). The architectures should \hl{also} be carefully designed so that they do not incur side-channels from resource sharing~(\id{AV-3-b}). Unfortunately, no commercial-grade accelerator architecture meets such requirements, despite the proposals from many academic works. We expect that practical accelerator hardware that is aware of confidential computing would be necessary for the widespread adoption of confidential ML computing.

%


\boxedpar{
Secure accelerator architectures retrofit existing accelerator designs (e.g., GPUs) for confidential computing. While they provide low overhead (17\%-33\%), only simulated research implementations \hl{exist} since they require hardware modifications to proprietary GPUs.  }

\label{subsec:accelerator-system}

\hdr{Hardware modifications to \hl{the} host system.} HIX~\cite{jang2019heterogeneous} secures GPU computations by introducing changes to the software and hardware stacks of the host system with a trusted processor. On the software side, the authors propose protecting the GPU driver within trusted enclaves. The key hardware modifications are added to the PCI-e interconnect and the \hl{\emph{memory management unit (MMU)}} to secure CPU-GPU communication against untrusted software. We observe that secure accelerator architectures such as Graviton~\cite{volos2018graviton} provide stronger security guarantees such as isolation among the GPU context inside GPU. Moreover, the communication between CPU and GPU in HIX is vulnerable to physical bus snooping, as traditional GPUs \hl{do} not support secure communication channel establishment.

\hdr{Secure I/O with weaker security guarantees.} 
On the other hand, there are attempts to deploy enclaves \hl{inside} virtual machines and let the hypervisor mediates \hl{the} secure use of accelerators~\cite{weiser2017sgxio,wang2020segive}. That is, they simply assume that the hypervisor is trustworthy and therefore capable of mediating secure connection between the enclaves and I/O devices. Although such methods do not require hardware changes to the accelerator or the host system, \hl{those} approaches \hl{adopt} a much weaker security model than that of SGX.

\boxedpar{
Approaches that introduce host-side modifications or hypervisor-based schemes for secure use of accelerators provide weaker security guarantees than secure accelerator architectures.
}


\section{\hl{ARM TrustZone and Computation in the Edge}}
\label{subsec:tz-side-channel}
\hl{While the majority of literature that proposes attacks and defenses on ML computations assume cloud computing scenarios where x86 processors are prevalent, ARM TrustZone has also been leveraged for confidential ML computations in mobile and edge devices}~\cite{mo2020darknetz,vannostrand2019confidential,schlogl2020ennclave,bayerl2020offline,sun2021shadownet,mo2021ppfl}. \hl{TZ has been employed for protecting ML models deployed on the edge and mobile devices for inference services}~\cite{mo2020darknetz,bayerl2020offline,sun2021shadownet}\hl{. Also, the federated learning scenario that we discuss in }\cref{subsec:fl}\hl{, necessitate the TZ-based training in mobile devices.}       

\hl{A survey (SoK) paper}~\cite{sok-tz}\hl{ provides a comprehensive overview of the currently known security vulnerabilities on TZ and its applications. Besides vendor-specific and implementation-specific vulnerabilities, there are microarchitectural attacks on the confidentiality of TZ-protected computation.} 

\hdr{\hl{Microarchitectural side-channels of TrustZone.}} \hl{In TZ-enabled CPUs, the cache lines are extended with a \emph{non-secure (NS)} bit that segregates cache usage of applications from the normal world and secure world.
However, programs from the two worlds have equal rights to content for the cache lines, which creates cross-world side-channels exploited by several works in the literature. ARMageddon}~\cite{lipp2016armageddon}\hl{ introduces a cross-core side-channel attack vector based on \emph{cache coherence mechanisms} between CPU cores to extract timing information from victim applications. On the other hand, using the \emph{cache contention side-channel} between the normal and secure worlds, several works~(e.g., TruSpy and TruSense) can extract sensitive information from TZ-protected computations}~\cite{zhang2016truspy,zhang2018trusense}.\hl{
Obtaining unprivileged timing sources is another challenge addressed by several attacks, as the \emph{performance counters}, commonly employed for precise timing information, are often inaccessible from the userspace. Several works utilize alternative sources of timing~(e.g., system calls and POSIX functions) to allows the side-channel attack to be performed by normal unprivileged applications}~\cite{lipp2016armageddon,zhang2018trusense,zhang2016truspy}. \hl{Prime+Count}~\cite{cho2018prime}\hl{ is a cache attack that assumes the adversaries can control applications from the secure world. The attack exploits the \emph{performance monitor unit (PMU)} to build a \emph{covert channel} that extorts data from the secure world to the normal world. Apart from cache side-channels, the shared usage of other microarchitectural features such as the \emph{branch target buffer (BTB)} is also exploited to leak sensitive information from confidential computations}~\cite{ryan2019hardwarebacked}. \hl{Nevertheless, the implications of TrustZone side-channels on confidential ML computation are underexplored, and we discuss it further in}~\cref{sec:opp-edge} \hl{for this reason.}


%% file: tables/attack-vectors.tex

\begin{table*}[ht]
    \centering
    \newcolumntype{Y}{>{\centering\arraybackslash}X}
    
 	\begin{threeparttable}
    \begin{tabularx}{\textwidth}{Ycp{4.5cm}Yp{4.5cm}}
        \toprule
        \thead{AV Groups} &\thead{Known Attacks\\Vectors} &  \thead{Description} & \thead{Known Attacks\\on ML} & \thead{Mitigation} \\
         \midrule
        \multirow{3}{\hsize}{\centering Untrusted SW} & \textbf{\texttt{AV-1-a}} & Unauthorized SW access to enclave memory  
        & - 
        & Mitigated by SGX 
        
        \\ 
        \noalign{\vskip 0.3em}    
        &\textbf{\texttt{AV-1-b}}  & Control\hl{led}-channels interactions with the untrusted OS (e.g., page faults, interrupts, system calls, etc.) & - & Oblivious ML  algorithms~\cite{grover2019privado,ohrimenko2016oblivious}\\
        \noalign{\vskip 0.3em}    
        & \textbf{\texttt{AV-1-c}}  & Microarchitectural Side-channels (e.g., cache, LFB, ... ) & \cite{yan2020cache,liu2020ganred,grover2019privado} & Oblivious ML  algorithms~\cite{grover2019privado,ohrimenko2016oblivious}\\
        \midrule
         \multirow{2}{\hsize}{\centering Physical Access} & \id{AV-2-a} & Unauthorized DMA access / Cold boot attacks on enclave & - & Mitigated by SGX \\
        \noalign{\vskip 0.3em}    
        &\id{AV-2-b}  & Memory bus side-channels & \cite{hua2018reverse} & Oblivious ML algorithms~\cite{ohrimenko2016oblivious,grover2019privado}\\ 
        \noalign{\vskip 0.3em}    

        &\id{AV-2-c}   & Power and EM side-channels & \cite{batina2019csi,yu2020deepem} & - \\ 
        \midrule
        \multirow{3}{\hsize}{\centering Accelerator} &\textbf{\texttt{AV-3-a}} &   Insecure I/O channel & \cite{zhu2021hermes} & 
        \makecell[l]{Blinding and verification schemes~\cite{tramer2019slalom,asvadishirehjini2020goat,sun2021shadownet,narra2019privacypreserving}, \\ secure accelerator architectures~\cite{volos2018graviton,hunt2020telekine}, \\ system support for secure I/O~\cite{jang2019heterogeneous,wang2020segive,zhu2019enabling,weiser2017sgxio}}
        \\  
        \noalign{\vskip 0.3em}    
        &\textbf{\texttt{AV-3-b}} & Side-channels in sharing GPU & \cite{jha2020deeppeep,wei2020leaky} & - \\ 
    
        \bottomrule

    \end{tabularx}
     \end{threeparttable}
    \caption{Attack Vectors on TEE-based confidential ML computation in \hl{the} cloud and their description. \hl{Related works on known attacks on ML computations using the attack vectors and their mitigation techniques are also included.}}
    \label{tab:attack-vectors}
\end{table*}

%% file: sections/multi-party.tex
\section{\hl{Satisfying} Multi-party Security Requirements}
\label{sec:multi-party}

In this section, we present our study on the existing literature that sought to satisfy varying security requirements in different TEE-based multi-party computation scenarios. We found that there can be multiple security goals in a single scenario, depending on the point of view of different entities. \autoref{tab:ai_scenario} illustrates the scenarios that we identified as multi-party ML computation, classified using our definitions from \autoref{tab:taxonomy}, \hl{and the security requirements from different point of views (e.g., service user versus service provider)}. The term \emph{multi-party computation} often refers to a subfield in cryptography. However, we use the term multi-party computation or multi-party ML computation to \hl{strictly} refer to \emph{TEE-based} approaches in this work. 

\input{tables/scenarios}

ML computation scenarios where there are more than one party involved have been extensively explored in the existing works. \hl{In addition to protecting computations with TEEs from the untrusted cloud,} the entities (\id{E-1}, \id{E-2}, and \id{E-3}) that contribute their data (\id{P-1}), ML model (\id{P-2}), or ML programs (\id{P-3}) are different parties and often mutually distrusting. As diverse stakeholders join and ownership of \hl{the} assets \hl{are} subdivided in the \hl{multi-party computation scenarios}, the security requirements for confidential ML applications \hl{have higher} standard \hl{than} a simple offloading scenario. More participants lead to increased \hl{communication complexity}, widen\hl{ed} boundary of communication, and larger attack surface. In contrast, data owners and model owners (or the service provider\hl{s}) desire \hl{the} privacy of their asset\hl{s} regardless of the situation. 

Most TEE-based \gls{mpml} schemes rely on the \emph{remote attestation} capability \hl{of TEE-enabled processors} to bridge the trust gap between distrusting parties. \hl{Remote attestation} allows remote parties to verify the integrity of code and data of a hosted enclave. Through remote attestation, each party can verify the code executing on the shared enclave before sending the protected asset to it. Ohrimenko et al.~\cite{ohrimenko2016oblivious} are the first to propose a TEE-powered privacy-preserving \gls{mpml} system, in which multiple data-contributing parties employ a trusted enclave hosted by a cloud provider to train a shared ML model while keeping their data secret from each other. The evaluation shows that the TEE-based approaches have a competitive execution time compared to cryptographic-based approaches for \gls{mpml}.

\Figure[t]()[width=0.95\columnwidth]{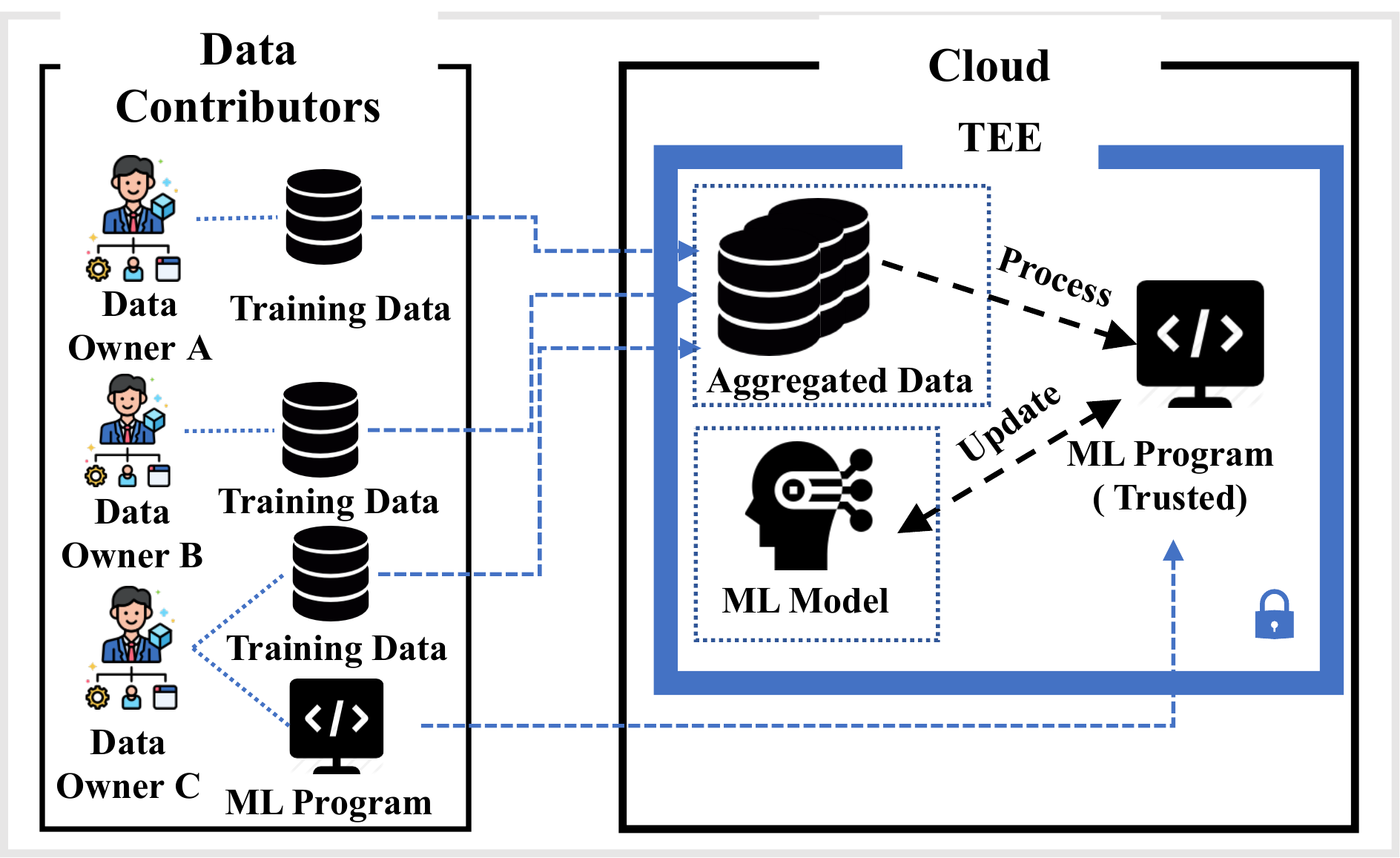}
  { Scenario \id{S-1} \hl{Collaborative ML. Mutually distrusting Data contributors} (\id{E-1}) \hl{contribute their dat}a \id{P-1} \hl{to collectively train a model that is to be shared among them. The TEE-based ML computation system must ensure the confidentiality of the data for each contributor}.  \label{fig:collab}} 

\subsection{Collaborative ML and Multi-party MLaaS}
\label{subsec:cloud-multiparty}

\subsubsection{Scenario \id{S-1}: Collaborative ML \hl{(multiple data contributors, one shared model)}} \hl{\emph{Collaborative ML} involves mutually distrusting data contributors who seek to train a co-owned ML model without revealing their data}~\cite{ohrimenko2016oblivious,hynes2018efficient}. \hl{All mutually distrusting parties} (\id{E-1-a}) \hl{have identical security requirements on their data}~(\id{P-1})\hl{, that is, data is not leaked to other contributors and the untrusted computing platform}. \hl{The scenario is often discussed along with the traditional MPC}~\cite{multiparty-unconditional,mental-game,yao1,yao2} \hl{that seeks to provide cryptographic primitives for computing a result of common interest without revealing the data from multiple contributors.} \hl{While the motivations are vastly similar, we only discuss TEE-based implementations works that focus on ML computations.}

\hdr{Trustworthy collaborative ML.} \autoref{fig:collab} illustrates the common approach \hl{to build a} trustworthy collaborative ML scheme \hl{using TEEs}. The same approach is used by Ohrimenko et al.~\cite{ohrimenko2016oblivious}. The proposed systems allow the contributing parties to verify the ML model and program via remote attestation before sending data. The system's goal is to enforce trustworthy ML computation such that no party can learn of the other party's contributed data directly or indirectly, and only allow the final ML model to be visible to the contributors. The key idea is to allow the contributors to verify that only privacy-preserving ML training algorithms can be performed inside the enclave. The proposed system employs an enclave trusted by all data owners to perform machine learning operations on private data. Moreover, the authors provide data-oblivious machine learning algorithms as an additional feature to prevent memory access side-channels. \hl{Later}, Myelin~\cite{hynes2018efficient} improve\hl{s} upon \hl{the system of}~\cite{ohrimenko2016oblivious} by applying differential privacy algorithms to the output of the trained model, \hl{enhancing} the data privacy of the resulting model.

\boxedpar{TEE-based collaborative ML computation enforces only privacy-preserving algorithms that all contributors agree upon to be used.}

\Figure[t]()[width=0.95\columnwidth]{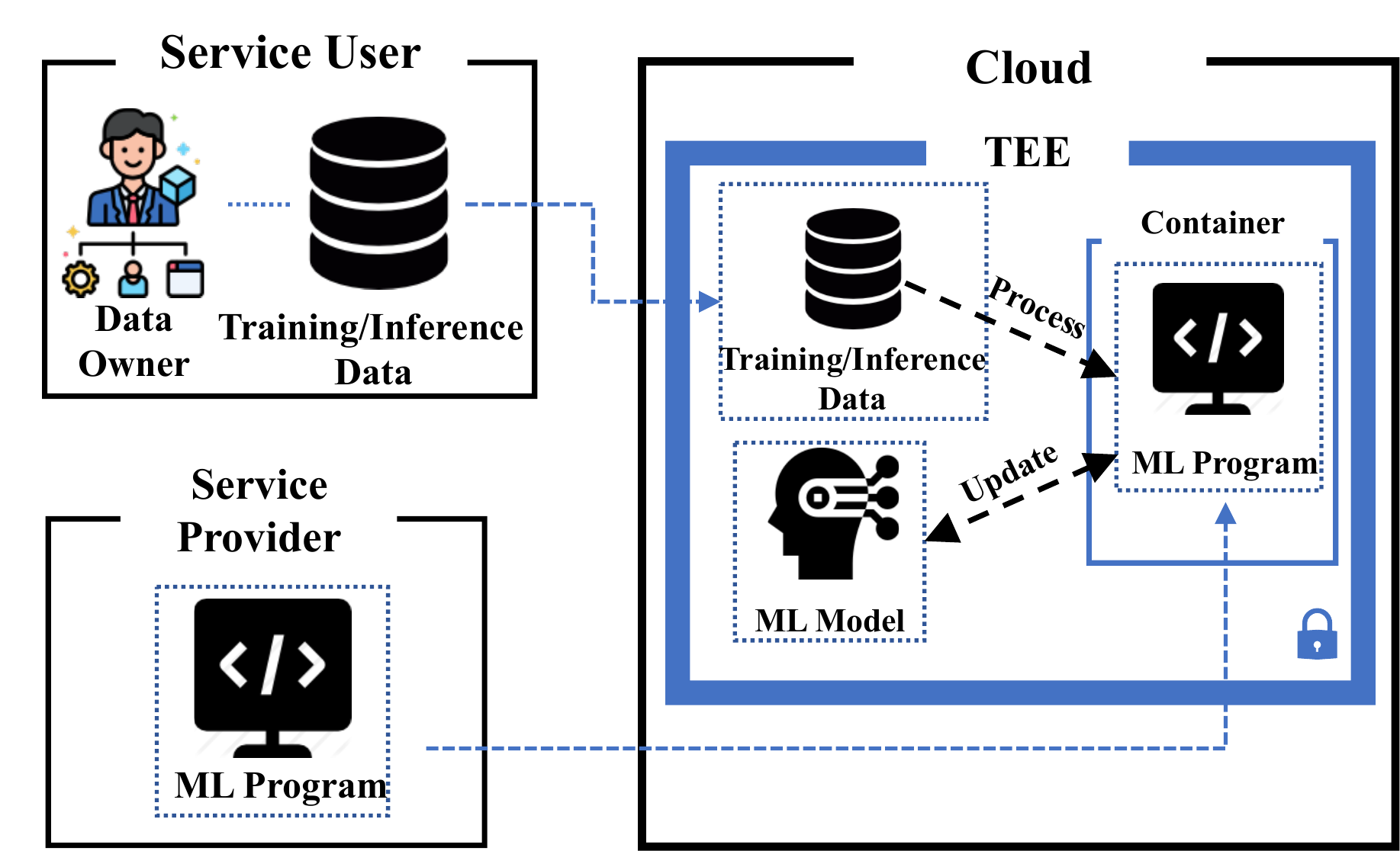}{\hl{Scenario} \id{S-2} \hl{multi-party MLaaS. MLaaS service provider}(\id{E-2}) \hl{owns the ML model} (\id{P-2}) \hl{and ML program} (\id{P-3}) \hl{to service users} (\id{E-1}) \hl{who input their data} (\id{P-1}). \hl{Since data confidentiality depends on the ML model and how an ML program trains or inferences using the data, a conflict in security requirement arises.} \label{fig:multi-party}}

\subsubsection{Scenario \id{S-2}: Multi-party \hl{Machine Learning as a Service} \hl{(Different data contributors and model owners)}}

\hl{We use the term \emph{multi-party machine learning as a service (MPMLaaS)} to refer to multi-party computation scenarios in which the service provider provides a service based on the ML model and ML program, and allows the users to process their data using the service.} \autoref{fig:multi-party}\hl{ illustrates the approach for satisfying the security requirements commonly used in the literature}~\cite{hunt2016ryoan,hunt2018chiron,ozga2021perun}\hl{. This scenario differs from }\id{\hl{S-1}}\hl{, in which the mutually distrusting data contributors had the same security requirements; Not only that the security requirements from the service provider and the service users differ, they are also in conflict}~\cite{hunt2018chiron, hunt2016ryoan}.


\hdr{Conflict in security requirements.} \hl{We see that simultaneously satisfying the confidentiality of ML models }\hl{(}\id{\hl{P-2}}\hl{)} \hl{and programs} \hl{(}\id{\hl{P-3}}\hl{)} \hl{from the service provider and service user data }\hl{(}\id{\hl{P-1}}\hl{)}\hl{ results in a conflict}~\cite{hunt2018chiron,hunt2016ryoan}. \hl{Assume that the service provider does not reveal the contents of the ML model and ML program used in her SGX-protected service to the service users. The service user cannot inspect the program's logic and thus cannot be assured that the service would respect their privacy. It should be noted that the conventional SGX-based secure cloud computing methods only verifies the integrity and authenticity of the SGX enclave itself and the in-enclave program. Hence, several works have proposed methods to resolve the issue.}


\hdr{Data flow control with SFI and containers.} A number of works have leveraged \emph{Software Fault Isolation (SFI)} to provide mitigation to the issue. The general approach is to employ SFI to SGX-based MPMLaaS or SaaS frameworks such that the data flow control is enforced according to a given policy, thereby ensuring user data confidentiality without revealing program code. Ryoan~\cite{hunt2016ryoan} creates SGX-protected sandbox instances using Native Client~\cite{yee2009native} to enforce \hl{the} prevention of information leakage. Native Client provides SFI-based sandboxing that confines the interaction between the host and the client (e.g., through system calls) from inside the sandboxes to the host system. \hl{The} technique has been adopted to complement SGX's guarantees \hl{with capabilities to isolate data processing modules, creating two-way sandboxes}~\cite{ozga2021perun,hunt2016ryoan,hunt2018chiron}. \hl{The} later work, Chiron~\cite{hunt2018chiron}\hl{,} bridges the conflict in confidential ML computation by employing Ryoan containers; it ensures that an untrusted \hl{program cannot leak data} (from \hl{the} data owner's (\id{\hl{E-1}}) perspective) and maintain the confidentiality of ML at the same time. Finally, Perun~\cite{ozga2021perun} further generalizes \hl{MPMLaaS} parties into \emph{stakeholders} and employs a trusted \emph{security policy manager} enclave to enforce the security policies of each stakeholder and to provision \hl{the} secret between stakeholders' enclaves. 



\boxedpar{Applying SFI-based sandboxing for data leakage prevention can ensure ML program/model confidentiality and data protection simultaneously.}

\subsection{Secure On-device ML and Federated Learning}
\label{subsec:fl}

\subsubsection{Scenario \id{\hl{S-3}}: On-device ML} 
\Figure[b]()[width=0.95\columnwidth]{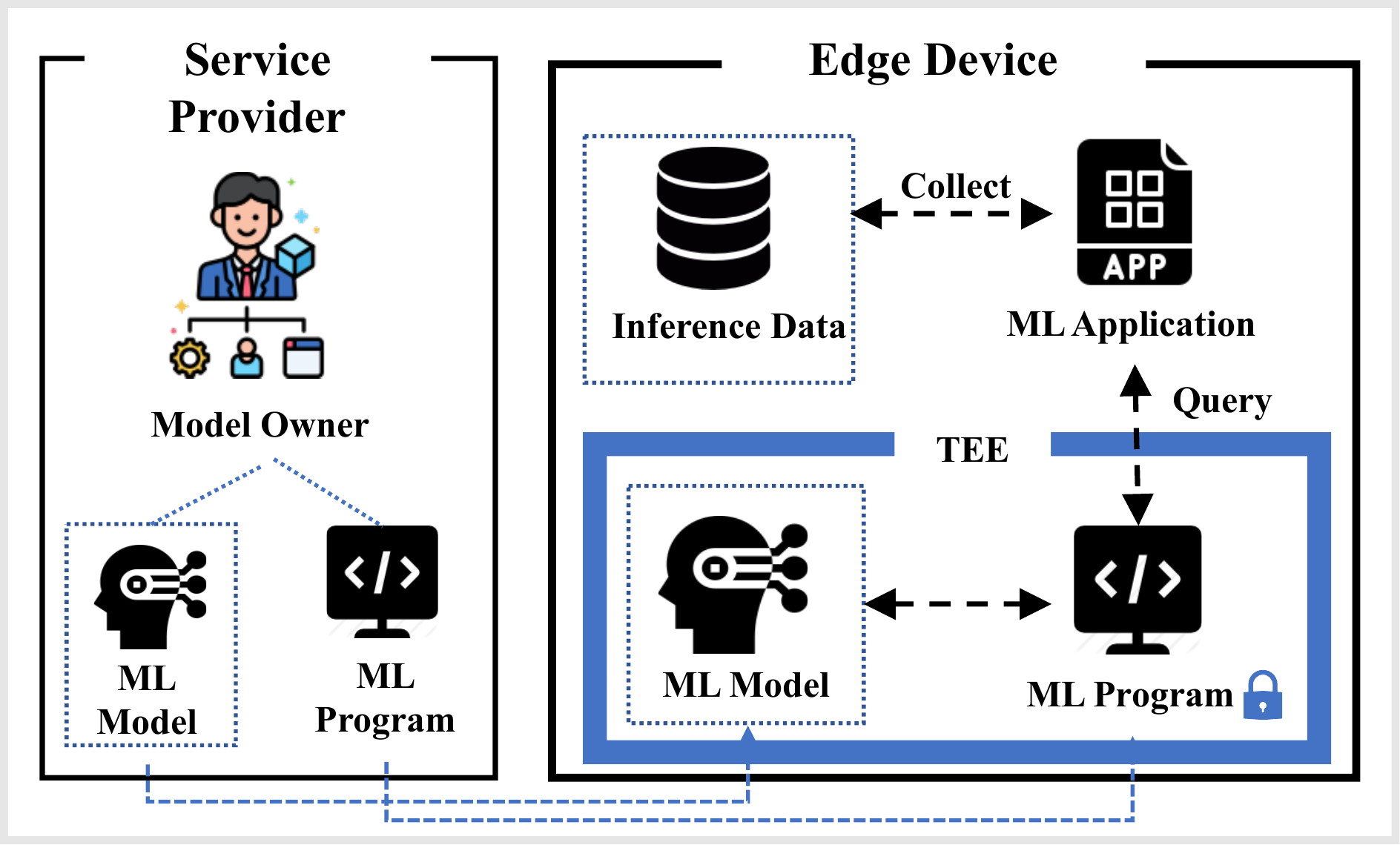}
  {\hl{Scenario} \id{\hl{S-3}} \hl{On-device ML. For responsive inference-based services, ML model owner} (\id{\hl{E2}}) \hl{may decide to deploy her ML Model} (\id{\hl{P-2}}) \hl{to user devices. In such cases, the user} (\id{\hl{E-1}}) \hl{may attempt to extract information from the deployed model or steal the model in its entirety.}  \label{fig:on-device}}
  
Deploying ML models to the edge or mobile devices \hl{has} become commonplace progressively. For instance, real-time inferences may be required for autonomous vehicles. Smartphone applications today often include an ML model that performs on-device inferences. The key security requirements are (1) to protect the confidentiality of ML models (\id{\hl{P-2}}) that can be the intellectual property of the service provider against curious or malicious device owner who attempts to reverse engineer the ML model and (2) to prevent leakage or abuse of on-device user data (\id{\hl{P-1}}) by distributed ML service. The work of Sun et al. \cite{sun2021mind} performs a large-scale study on mobile applications \hl{that} employ on-device ML. The authors point out that $41\%$ of the analyzed ML apps do not protect the ML model, and $66\%$ of ML apps that attempted to secure their ML model adopted insufficient protection. This particular work confirms the current state of confidential ML computation in the edge and mobile devices straightforwardly; demand for confidentiality in on-device ML is on the rise, and research opportunities lie on this.

\hdr{Protecting ML models in mobile TEEs.} \autoref{fig:on-device} illustrates the simplified approach for utilizing TEEs \hl{to protect ML models on} edge devices, \hl{which is to encrypt the ML model and only store it in plaintext inside of TEE-protected memories.} MLCapsule~\cite{hanzlik2021mlcapsule} is one of the preliminary works that allow a service provider to deploy an inference service on a user's device provides the same security guarantees and same levels of control over the model~(\hl{by} the model owner) as typical server-side execution. However, it uses SGX, a mostly cloud-based TEE, for the implementation and evaluation. \hl{The} majority of confidential on-device ML researches leverages ARM TrustZone, a TEE commonly found in the edge. Notably, Offline Model Guard~(OMG)~\cite{bayerl2020offline} leverages SANCTUARY~\cite{brasser2019sanctuary}, an SGX-like isolated execution environment powered by TZ to protect the computation of ML computation on the user's device. The authors also show how it can securely obtain the user's data from peripheral devices such as microphones and sensors, with one of TZ's features. While most proposed works on on-device ML focus on deploying pre-trained ML for inference, PrivAI~\cite{meurisch2020privacypreserving} allows the service users to \emph{personalize} proprietary ML models (i.e., update the model with the user's data) by deploying the training process inside the TEE on the user's device. 
Finally, instead of protecting the model's confidentiality, DarkneTZ~\cite{mo2020darknetz} hides \hl{only the} sensitive layers of a DNN to defend against the membership inference attack.
Apart from the aforementioned works, the remaining works on secure on-device ML address challenges incurred by the architectural limitations~\cite{schlogl2020ennclave,vannostrand2019confidential}, or secure usage of GPU~\cite{sun2021shadownet}, which we already covered in the other sections~(\cref{sec:engineering-challenges}). 

\boxedpar{Mobile ML \hl{applications} often employ insufficient protection for the model. TEEs \hl{are} required to securely protect \hl{the} ML model on the user's device.}

\subsubsection{Scenario \id{\hl{S-4}}: Federated Learning}

\Figure[t]()[width=0.95\columnwidth]{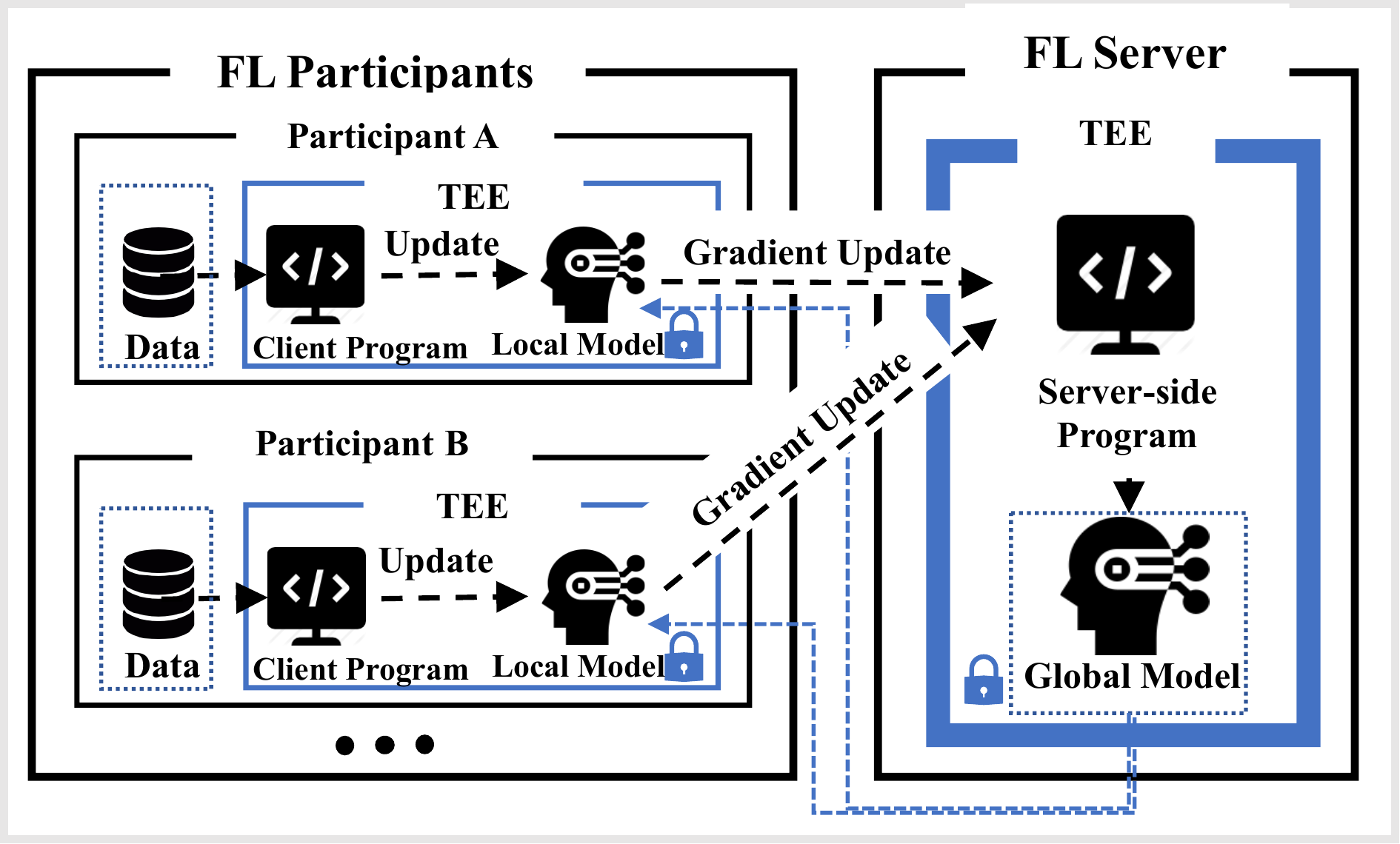}
   {Scenario \id{\hl{S-4}} \hl{federated learning. Server-side TEEs and client-side TEEs are often cooperate in federated learning scenarios. Service provider wants to enforce correct execution of training on user devices while users desire the confidentiality of the aggregated results.} \label{fig:fl}}
\emph{Federated learning (FL)} is a distributed ML training method that exploits \hl{the} parallelism of multiple machines to train a \emph{global model}. It allows \emph{FL participants} to train \emph{local models} using their data\hl{, and} collects only the \hl{participant's} \emph{gradient updates}. 
FL excludes the data collection step of traditional ML workflows, so it attains an advantage on privacy \hl{over} others ML paradigms~\cite{kairouz2021advances,mcmahan2017communicationefficient}. TEE can further fortify FL, which is already privacy-preserving to a certain degree without TEEs. 

FL introduces a two-fold security requirement for \hl{each of} the participating \hl{party}. From \hl{the} FL participant(client) perspective, the service provider can be dishonest and back-channel user data (\id{\hl{P-1}}). The device owner, who performs training on his or her device, also wants confidentiality of the collected gradient from training.
On the other hand, the service provider may want the deployed ML model/program to be protected (\id{\hl{P-2}}, \id{\hl{P-3}}) or trained without malicious alterations from service users and attackers with access to the device. 

\input{tables/cloud-comparison-bak}

\hdr{Server-side secure gradient aggregation.}
Although only the gradient obtained by the clients \hl{is} sent to the server \hl{in the FL paradigm}, several works demonstrated that these updates \hl{might} leak unintentional private data~\cite{melis2019exploiting, zhu2019deep, geiping2020inverting}. The secure aggregation algorithm proposed by Bonawitz et al.~\cite{bonawitz2016practical} solves this issue by leveraging secure MPC. However, secure aggregation cannot provide full coverage. For example, there is no guarantee that the server correctly implements the protocol. Multiple works~\cite{mo2021ppfl,lin2021esmfl,zhang2021shufflefl} perform the gradient aggregation process inside a server-side TEE to protect the gradient of participants from adversaries. The systems use remote attestation to verify the authenticity of FL servers.

\hdr{Integrity of client-side computation. } Other works employ client-side TEEs to ensure that client-side FL computation \hl{is} executed correctly. A malicious FL client could impair the integrity of the global model by sending erroneous gradient updates to the FL server ~\cite{tolpegin2020data}. Zhang et al. proposed TrustFL~\cite{zhang2020enabling} for client-side computation integrity. In their scenario, honest servers do not trust their clients~(or data owners). TrustFL leverages the TEE of clients to identify the training processes and ensure the integrity of results. SEAR~\cite{zhao2021sear} employs TEEs to solve the Byzantine Generals Problem that arises in FL due to malicious participants.

\hdr{End-to-end protection. } \autoref{fig:fl} demonstrates a design for \hl{an} FL system protected by TEEs on both the participants' devices and the server. A similar approach is employed by PPFL \cite{mo2021ppfl}\hl{,} which utilizes TEEs on both the mobile device and \hl{the} server to conceal training and data aggregation. On the device side, the entire training process happens inside a TEE \hl{to prevent tampering}. On the server\hl{-}side, SGX protects the data aggregation program \hl{from the untrusted cloud service provider}. Their work shows that the proposed system is robust to data reconstruction, property inference, and membership inference attacks.

\boxedpar{ Compared to on-device ML deployment, the necessity of a trusted gradient aggregator emerges in federated learning. Also, because of its decentralized nature, FL encounters a few known challenges of decentralization.}

\subsection{Summary}
\autoref{tab:multi-party} shows the summary of works that we discussed \hl{in this section}. The table lists works that fall into our defined multi-party ML scenarios, the TEEs they utilized, protected assets~(i.e., data, ML model, and ML program), and the computation performed~(i.e., training or inference). 

\hdr{Cloud-based ML deployment scenarios. } \id{\hl{S-1}} and \id{\hl{S-2}} \hl{commonly seeks to shield the computation from the untrusted cloud infrastructure using server TEEs. Also, they both assume mutually distrusting service users and ensures the data confidentiality to each user}~\cite{ohrimenko2016oblivious, hynes2018efficient, hunt2018chiron, hunt2016ryoan}. \hl{That is, there are multiple parties that have the same security requirements in case of} \id{\hl{S-1}}.

\id{\hl{S-2}} \hl{differs from} \id{\hl{S-1}} \hl{in that the party that owns the ML model and ML program that performs training or inferences, and the service users are also mutually distrusting. Hence, there exists two point of views} (\id{\hl{E-1}} and \id{\hl{E-2}}) \hl{with differing security requirements as shown in }\autoref{tab:ai_scenario}. \hl{For this reason, works such as Chiron}~\cite{hunt2018chiron} \hl{proposed SFI-based data flow assurence to resolve the conflict.}

\hdr{Edge-based ML deployment scenarios. } \hl{Both} \id{\hl{S-3}} \hl{and} \id{\hl{S-4}} \hl{involve deploying ML computations to the users' devices. In case of} \id{\hl{S-3}}, \hl{the service provider desires the confidentiality of the ML model deployed in user devices. On the other hand, since the inference using ML model is performed locally, user's data and inference results can be contained inside the device. Hence, Many existing works proposed TEE-based solution for secure on-device ML model protection and safeguarded inferences}~\cite{vannostrand2019confidential,schlogl2020ennclave,bayerl2020offline, meurisch2020privacypreserving, mo2020darknetz, hanzlik2021mlcapsule, sun2021shadownet}.

\hl{For federated learning  }(\id{\hl{S-4}})\hl{, the ML models are often not considered confidential as they are to be trained in the user devices. Instead, the service providers need to ensure correct execution (i.e., integrity) of the client-side training. On the other hand, the clients want confidentiality of the aggregated gradients from training.} 
%



%% file: tables/scenarios.tex
\label{subsec:scenarios}
\begin{table*}[ht]
    \centering
	\newcommand*\feature[1]{\ifcase#1 \Circle\or\LEFTcircle\or\CIRCLE\fi}
    \newcommand\comparison[1]{%
        \setsepchar{,}%
        \greadlist*\args{#1}%
        \feature{\args[1]}%
        &\feature{\args[2]}%
        &\feature{\args[3]}%
    }
    \newcolumntype{Y}{>{\hsize=1.2\hsize\centering\arraybackslash}X}
    \newcolumntype{M}{>{\hsize=0.3\hsize\centering\arraybackslash}X}
    \newcolumntype{S}{>{\hsize=0.2\hsize\centering\arraybackslash}X}
    \begin{tabularx}{\textwidth}{cp{2.5cm}p{3.2cm}*{3}{S}X}
       \toprule
        &
        &
        & \multicolumn{3}{c}{\thead{Protected Assets}}
        &
       \\
      \cmidrule(lr){4-6}
         \thead{ID} 
         & \thead{Scenario} 
         & \thead{Point of View} 
         & Data (\id{P-1}) & ML Model (\id{P-2}) & ML Program (\id{P-3})
         & \thead{Attacker}\\
         \midrule
        \id{S-1} & Collaborative ML 
        & Data contributors~(\id{E-1-a})
        & \comparison{2,0,0,2,2,0,0,2}
        & \makecell{Mutually distrusting contributors~(\id{E-1-a}') \\ Untrusted computing platform~(\id{E-3})}
        \\ 
        \midrule
        \multirow{3}*{\id{S-2}} &\multirow{3}*{Multi-party MLaaS}          
        & Service user~(\id{E-1-b})
        &  \comparison{2,0,0,2,2,0,2,0}
        & \makecell{Untrusted service provider~(\id{E-2}) \\ Untrusted computing platform~(\id{E-3})} \\
        \cmidrule(lr){3-7}
        & & Service provider~(\id{E-2})
        &  \comparison{0,2,2,2,2,0,2,2}
        & \makecell{Untrusted service users~(\id{E-1-b}) \\ Untrusted computing platform~(\id{E-3})}
        \\ 
        \midrule
        
       \multirow{3}*{\id{S-3}} & \multirow{3}*{On-device ML}          
        & Service user~(\id{E-1-\{a,b\}})
        &  \comparison{2,0,0,0,0,0,2,0} 
        & \makecell{Untrusted service provider~(\id{E-2})}
        \\ 
        \cmidrule(lr){3-7}
        && Service provider~(\id{E-2})
        &  \comparison{0,2,2,0,0,2,2,2} 
        & \makecell{Untrusted service users~(\id{E-1-\{a,b\}}) \\ Untrusted computing platform~(\id{E-3})}
        \\ 
        \midrule
        
        \multirow{3}*{\id{S-4}} &\multirow{3}*{Federated Learning}          
        & FL participant~(\id{E-1-a})
        &  \comparison{2,0,0,2,2,0,2,2} 
        & \makecell{Mutually distrusting FL participants~(\id{E-1-a}') \\ Untrusted service provider~(\id{E-2}) \\ Untrusted computing platform~(\id{E-3})}
        \\ 
        \cmidrule(lr){3-7}
        && FL server~(\id{E-2})
        &  \comparison{0,2,2,2,2,2,2,2} 
        & \makecell{Untrusted participants~(\id{E-1-a}) \\ Untrusted computing platform~(\id{E-3})}
        \\

       \bottomrule
    \end{tabularx}
    \caption{\hl{Multi-party ML scenarios by protected assets and attacker model. Different security requirements from different participating parties are the unique characteristic of multi-party ML computation scenarios. The table shows  generalized protected assets vs. attackers model for each point of view in multi-party ML scenarios.} }
    \label{tab:ai_scenario}
\end{table*}

%% file: tables/cloud-comparison-bak.tex
\begin{table*}[t]
	\centering
	\footnotesize
    \newcolumntype{Y}{>{\centering\arraybackslash}X}
	\newcommand*\rot[1]{\hbox to1em{\hss\rotatebox[origin=lb]{90}{\noindent#1}}}
	\newcommand*\feature[1]{\ifcase#1 \Circle\or\LEFTcircle\or\CIRCLE\or\CIRCLE\textsuperscript{\dag}\or-\or\CIRCLE\textsuperscript{*}\fi}
	
	\newcommand*\fff[3]{\feature#1&\feature#2&\feature#3}
	\newcommand*\ff[2]{\feature#1&\feature#2}
	\newcommand*\f[1]{\feature#1}
	\makeatletter
	\newcommand*\ex[9]{#1\tnote{#2}&#3&%
		\fff#4&\fff#5&\f#6&\f#7&\fff#8&\fff#9
	}
	
    \newcommand*\work[4]{#1\tnote{#2}&#3&#4}
    \newcommand\comparison[1]{%
        \setsepchar{,}%
        \greadlist*\args{#1}%
        \feature{\args[1]}%
        &\feature{\args[2]}%
        &\feature{\args[3]}%
        &\feature{\args[4]}%
        &\feature{\args[5]}%
    }
	\makeatother
	\begin{threeparttable}
		\begin{tabularx}{\textwidth}{p{4cm}cc *{6}{Y}c*{4}{Y}}
			\toprule
			\multirowthead{3}[-2em]{Approach}  
			& \multirowthead{3}[-2em]{Year} 
			& \multirowthead{3}[-2em]{TEE} 
			& \multicolumn{3}{c}{\thead{Protected Asset}}  
			& \multicolumn{2}{c}{\thead{Computation Performed}}  
			\\ 
			\cmidrule(lr){4-6}\cmidrule(lr){7-8}
		    &	
		    & 
			& Data \hspace{3em} (\id{P-1})
			& ML Model (\id{P-2})
			& ML Program (\id{P-3})
			& Training 
			& Inference 
			\\
			\midrule
			\thead{\id{S-1}: Collaborative ML}\\   
			\cmidrule{1-1}  
			\work{Ohrimenko et al.~\cite{ohrimenko2016oblivious}}{}{2016}{SGX}                     &\comparison{2,2,0,2,0,0,0,0,0,0,0,0,0}\\
			\work{Myelin~\cite{hynes2018efficient}}{}{2018}{SGX}                                    &\comparison{2,2,0,2,2,0,0,0,0,0,0,0,0}\\
			\midrule
			\thead{\id{S-2}: Multi-party MLaaS}\\   
			\cmidrule{1-1}  
			\work{Chiron~\cite{hunt2018chiron}}{}{2018}{SGX}                                       &\comparison{2,2,2,2,2,0,0,0,0,0,0,0,0}\\
			\work{Citadel~\cite{zhang2021citadel}}{}{2021}{SGX}                                    &\comparison{2,2,2,2,2,0,0,0,0,0,0,0,0}\\
			\work{Perun~\cite{ozga2021perun}}{}{2021}{SGX}                                         &\comparison{2,2,2,2,2,0,0,0,0,0,0,0,0}\\
			\midrule
			\thead{\id{S-3}: On-device ML}\\   
			\cmidrule{1-1}  
			\work{Vannostrand et al.~\cite{vannostrand2019confidential}}{}{2019}{TZ}          &\comparison{4,2,0,0,2,0,0,0,0,0,0,0,0,0,0} \\   
			\work{eNNclave~\cite{schlogl2020ennclave}}{}{2020}{TZ}                                &\comparison{4,2,0,0,2,0,0,0,0,0,0,0,0,0,0} \\ 
			\work{Offline Model Guard~\cite{bayerl2020offline}}{}{2020}{TZ}                       &\comparison{4,2,0,0,2,0,0,0,0,0,0,0,0,0,0} \\ 
			\work{PrivAI~\cite{meurisch2020privacypreserving}}{}{2020}{SGX}                       &\comparison{4,2,0,2,2,0,0,0,0,0,0,0,0,0,0} \\ 
			\work{DarkneTZ~\cite{mo2020darknetz}}{}{2020}{TZ}                           &\comparison{5,1,0,0,2,0,0,0,0,0,0,0,0,0,0} \\
			\work{MLCapsule~\cite{hanzlik2021mlcapsule}}{}{2021}{SGX}                             &\comparison{4,2,0,0,2,0,0,0,0,0,0,0,0,0,0} \\ 
			\work{ShadowNet~\cite{sun2021shadownet}}{}{2021}{TZ }                                 &\comparison{4,2,0,0,2,0,0,0,0,0,0,0,0,0,0} \\ 
			\midrule
			\thead{\id{S-4}: Federated Learning}\\   
			\cmidrule{1-1}  
			\work{TrustFL~\cite{zhang2020enabling}}{}{2020}{SGX(\textbf{S},\textbf{C})}                                  &\comparison{3,0,0,2,0,0,0,0,0,0,0,0,0,0,0} \\
			\work{PPFL~\cite{mo2021ppfl}}{}{2021}{SGX(\textbf{S}),TZ(\textbf{C})}                                           &\comparison{3,2,0,2,0,0,0,0,0,0,0,0,0,0,0} \\
			\work{SEAR~\cite{zhao2021sear}}{}{2021}{SGX(\textbf{S})}                                          &\comparison{3,0,0,2,0,0,0,0,0,0,0,0,0,0,0} \\
			\work{ShuffleFL~\cite{zhang2021shufflefl}}{}{2021}{SGX(\textbf{S},\textbf{C})}                              &\comparison{3,0,0,2,0,0,0,0,0,0,0,0,0,0,0} \\
			\work{ESMFL~\cite{lin2021esmfl}}{}{2021}{SGX(\textbf{S})}                              &\comparison{3,0,0,2,0,0,0,0,0,0,0,0,0,0,0} \\
			\bottomrule
		\end{tabularx}
		\begin{tablenotes}
			\item \hfil$\feature2=\text{provides property}$; $\feature1=\text{partially provides property}$;
			$\text{\feature0}=\text{does not provide property};$ 
		    \item \hfil \textbf{C} $= \text{client-side TEE}$; \textbf{S}$= \text{server-side TEE}$;
		    \item \hfil - Not applicable; \textsuperscript{*} protected against membership inference; \textsuperscript{\dag} \hl{In FL, the local gradients obtained from data are protected instead of data}.
		\end{tablenotes}    
		\caption{Classification of existing works based on multi-party scenarios. \hl{Many works focus on a specific security requirement among multiple security requirements that exist in multi-party ML computation scenarios. The table consolidates the existing ad-hoc efforts to provide an overview by classifying existing works based on protected assets and type of computation performed}.}
		\label{tab:multi-party}
	\end{threeparttable}
\end{table*}


%% file: sections/engi-challenge.tex
\section{Engineering Challenges in Building ML computations inside TEE}
In this section, we discuss the engineering challenges in building systems for confidential ML computation. Besides security issues, retrofitting existing software and hardware to achieve confidential ML has also been an important issue in the field of systems security.

\label{sec:engineering-challenges}
\subsection{Memory limitations of TEEs} 
\label{subsec:memory-limit}

\hdr{Limited memory capacity in SGX. }Various works that apply confidential ML to the cloud struggles with the inherent memory limitation in SGX. As the size of data used for training is ever-growing, the memory capacity limitation of SGX may hinder the adoption of the technology. Hence, many works proposed optimizations that allow data-intensive computations to run in SGX enclaves.

The current version of Intel SGX (v1.0) has a hard limit of 96MB (excluding the 32MB reserved memory)~\cite{costan2016intel}. This memory limitation also leads to additional performance overhead from swapping memory pages from EPC memory to non-EPC memory~\cite{taassori2018vault}. Due to this reason, several works evaluate their approach \hl{with the memory} limitation~\cite{priebe2018enclavedb}. Other works employ the simulation mode of SGX for evaluation, with the hope that future versions of SGX will support larger memory size~\cite{priebe2020sgxlkl}. 
Currently, some cloud providers provide virtual machines with Intel SGX 2.0 that allow up to 1TB of EPC memory~\cite{microsoftazure,alibaba-cloud}. However, processors with Intel SGX 2.0 have yet to be widely available~\cite{occlum}, and their mechanisms are still unclear. We expect that the elimination of the memory limit would likely solve many of the technical challenges that we mention in this section.

\hdr{Memory capacity in ARM TrustZone.} ARM TrustZone, a TEE commonly deployed for mobile and edge ML computations, also suffers from memory limit issues. While Trustzone does not have a hardware-set limit, the amount of memory has to be decided for the normal world and secure world (Trustzone-side) during boot and cannot be changed afterward. Since most of the applications run in the normal world, and normal world performance directly impacts user experience, the secure world tends to have a small default memory limit.

\hdr{Splitting ML workload into batches.}
As \hl{the sizes of} ML models often far exceed the memory capacity of most TEEs, the most straightforward way to overcome the memory limitation is to split the ML models into smaller partitions and process them \hl{one by one} inside enclaves. The partitioning of most DNN models can be done without much effort, as DNN models are often subdivided into multiple layers. The remaining challenges are how to partition the models and what partitioning schemes offer the most security benefits. To this end, Vannostrand et al. \cite{vannostrand2019confidential} experimented with several partitioning schemes, namely \emph{layer-based partitioning}, \emph{sub-layer partitioning}\hl{,} and \emph{branched partitioning}. The authors found that each partitioning scheme is useful depending on the model size.

\hdr{Selective layer protection.} Several works aim to identify the minimal set of sensitive layers, e.g., layers that contain information of the input~\cite{mo2020darknetz,gu2020confidential,sun2021shadownet,schlogl2020ennclave}, and only protect those layers inside secure enclaves. Notably, Gu et al.~\cite{gu2020confidential} propose\hl{d} a ternary partitioning scheme, where the ML model is partitioned into \hl{a} sensitive \hl{section} that consists of the input layers and output layers and the remaining non-sensitive layers. Occlumency~\cite{lee2019occlumency}, on the other hand, assumes that the \gls{dnn} model does not require confidentiality and proposes techniques to load parts of the model into the enclave secure memory from insecure memory during inference. On edge, Mo et al.~\cite{mo2020darknetz} propose protecting only the last layer of a DNN inside TruztZone to thwart white-box membership inference attacks (MIA) effectively.


 

Moreover, a security problem arises from memory limitation. Malicious entities could observe the access pattern when data is moved in and out of an enclave's secure memory for computation. Previous works have shown that access patterns alone could leak sensitive information, even when the memory content is encrypted~\cite{islam2012access}. To solve the access pattern leakage, applying an oblivious computing algorithm has been suggested to guarantee the safeness of the model. Several works adapted ORAM schemes~\cite{sasy2018zerotrace,ahmad2018obliviate,ahmad2019obfuscuroa} or \hl{secure} hardware storage~\cite{oh2020trustore} to extend the memory limitation and eliminate the access pattern side-channel. Trustore~\cite{oh2020trustore} is a proposal that utilizes a PCI-e FPGA device to provide additional storage for SGX enclaves that also provides oblivious memory guarantees.

\hdr{Optimizing memory usage of ML model. } 
Several works reduce the ML model size through the use of TensorFlow Lite, a framework designed for a mobile device with limited capability~\cite{kunkel2019tensorscone,tf-trusted}. The framework applies model reduction through quantization to reduce the memory footprint of ML models. On the other hand, other works target the inefficiency of memory usage by enclave applications. In particular, Vessels~\cite{kim2020vessels} point out that the default scheme for memory usage in SGX employs a large allocation size and has low memory reusability. Based on the observation, the authors propose mechanisms to optimize the memory usage in ML computation that reduce up to $90\%$ of the overall memory footprint while also improving \hl{the execution time}.

\boxedpar{Partitioning ML models and optimizing the memory usage helps partially overcome the memory limitation of TEEs.}


\subsection{Performance overhead}

\hdr{TEE Transition overhead.} Use of TEEs requires additional performance overhead. A common source of performance in TEEs is the cost of transition. Just as the transition between kernel mode and user mode, the transition between the untrusted domain and TEE also adds overhead. However, the transition between the untrusted code and enclave is rather costly, as explored in the works that profiled the performance characteristics of Intel SGX~\cite{costan2016intel}. As ~\cite{zhao2016performance} provides, the number of cycles consumed for entering an enclave (\texttt{OCALL}) exiting (\texttt{ECALL}) is around 35 times that of an average system call. Generally, the frequency of the transition between the untrusted domain and TEE domain is directly proportional to performance overhead.

\hdr{Coping with limited memory.} TEEs often have memory limitations\hl{,} as mentioned above, and workarounds proposed in previous works (e.g., ~\cite{lee2019occlumency, gu2020confidential,narra2019privacypreserving}) introduce additional overhead. Since many such strategies for coping with limited memory involve splitting the workload into smaller batches, they also increase the transition overhead.

\boxedpar{The memory limitation of SGX is one of the predominant sources of performance overhead in SGX, especially for ML. Also, ML applications in SGX should minimize the number of TEE state transitioning.}

\subsection{Porting ML frameworks in TEEs} 


\hdr{Cost of porting ML programs to TEE\hl{.}} Porting \hl{complex programs} such as the ML frameworks (e.g., Tensorflow~\cite{tensorflow}, PyTorch~\cite{pytorch}) to use SGX is a daunting task for developers. The SGX SDK offers a set of build tools and APIs in low-level languages (C/C++). Many ML frameworks are complex and include components that are in higher-level languages (e.g., Python).

\hdr{Deploying ML frameworks inside TEEs.} Due to the high cost of porting complex ML frameworks, many existing works have proposed using \emph{software containers} such as Docker~\cite{merkeldocker} and \emph{Library OSes (LibOSes)}~\cite{tsai2017graphenesgx,baumann2014shielding,priebe2020sgxlkl,occlum} as alternative ways to deploy programs to cloud enclaves. Software containers \emph{virtualize} all interaction of a program with the containing system~(e.g., an SGX enclave), allowing the code to be portable across platforms. Such approaches allow ML frameworks to run in TEEs without much modification to their codebase. SCONE~\cite{arnautov2016scone} is a Linux container system that protects Docker containers~\cite{merkeldocker} with SGX enclaves. Later, TensorSCONE~\cite{kunkel2019tensorscone} introduces TensorFlow to SCONE containers, while secureTF~\cite{quoc2020securetf} introduces a secure distributed machine learning framework built upon TensorFlow and SCONE. Library OSes~\cite{baumann2014shielding,priebe2020sgxlkl,tsai2017graphenesgx,occlum} are types of containers that include all dependencies to a program, including an OS kernel emulation layer and modified standard libraries. LibOSes allows rapid and convenient deployment of programs, but the trusted code base inside an enclave significantly increases. The availability of a secure version of ML frameworks that are widely used allows cloud users to deploy secure ML computation on the cloud with ease. 





\boxedpar{Software containers allow unmodified ML programs to run inside \hl{SGX} but significantly increases the trusted \hl{codebase}.}


%

%% file: sections/opportunities.tex
\section{Future Directions}
\label{sec:research-opportunity}

Through our survey, we found issues that are underexplored or cannot be resolved with the discoveries in the \hl{existing} works. We provide our outlook on the possible future directions in this section.

\subsection{Confidential ML in Edge and End-point Devices}
\label{sec:opp-edge}

A predominant number of works that we reviewed contributed to SGX-based cloud ML computations. On the other hand, we observed that the number of works focused on ML computations using ARM TrustZone is relatively scarce. However, on-device ML computations have become more common~\cite{sun2021mind}, and works that explore confidential ML in the edge and end-point are on the rise. \hl{Consequentially,} a few works have discussed the \hl{security} issues \hl{exist in the scenario}~(e.g., ~\cite{bayerl2020offline,schlogl2020ennclave,mo2020darknetz}). 
\hl{Moreover, although the security vulnerabilities and side-channels of TrustZone are well-studied (systemized by Cerdeira et al.}~\cite{sok-tz}\hl{), little is known about their implications on confidential ML computations.}
We expect that more in-depth exploration of \hl{the issues in deploying confidential ML computations to the edge} will follow in the future.

\subsection{Secure Accelerator Architectures}
 An accelerator such as GPU plays a crucial role in achieving high performance in ML training and inference. However, there is no commercial-grade accelerator architecture currently available. All SGX-based confidential ML computation papers that we reviewed assume CPU-only computation for this reason. Graviton~\cite{volos2018graviton} proposes a GPU architecture that can ensure the confidentiality of the workload offloaded from Intel SGX. However, Graviton design includes hardware modifications to the GPU, which is not feasible due to proprietary architecture and firmware. Hence, the authors emulate hardware modifications to evaluate their design. SafeTPU~\cite{meracollantes2020safetpu} \hl{is another proposed secure accelerator architecture but} focuses on verifiable computation results and does not provide confidentiality. We expect that research towards secure acceleration of ML computations will be on the rise. For instance, various accelerators (e.g., \emph{neural processing units (NPUs)}) designed with confidential ML computing considerations will be of great contribution to the field.

\subsection{Confidential On-line Training.} Most works in secure on-device ML deployment only discuss \hl{the} ML model's confidentiality during inference. We found that secure \hl{updates }(e.g., \emph{personalizing} pre-trained ML models) \hl{are} rarely discussed in the existing works. Personalization of ML models can be useful in several real-life scenarios. For instance, ML models for self-driving cars would need to be adapted to the users' surrounding environment for better performance. We expect that designing a system that has to support inference and training at the same time will face extended attack surface\hl{s} and unique problems.

\subsection{Comprehensive Approach to Model Confidentiality}
\hl{This survey covered the existing works that discovered attacks on model confidentiality through in-system attacks, such as side-channels in the untrusted cloud. However, ML model confidentiality can also be undermined through API attacks that reconstruct the model based on the repeated inference results}~\cite{tramer2016stealing,jagielski2020high}. \hl{The attacker in the untrusted cloud who can launch side-channels attacks on an ML model in service can also make inferences to the service. However, the ramifications of such an attack that combines in-system attack vectors and API attacks have not been explored. We expect a comprehensive approach to model confidentiality that combinest the existing works that discuss threats in an ad-hoc manner could be of great value as future work.}